\documentclass[12pt]{article}

\pdfoutput=1

\usepackage[top=80pt,bottom=85pt,left=85pt,right=85pt]{geometry}
\usepackage{amssymb}
\usepackage{amsmath}
\usepackage{mathtools}
\usepackage{setspace}
\usepackage{accents}
\usepackage{comment}
\usepackage[english]{babel}
\usepackage{hyphenat}
\usepackage[shortcuts]{extdash}

\usepackage[utf8]{inputenc}
\usepackage{uniinput}
\usepackage{changepage,cite}
\usepackage[plain]{fancyref}

\usepackage{makecell} 
\usepackage{longtable} 
\usepackage[normalem]{ulem} 
\usepackage{transparent} 
\usepackage{bm} 

\usepackage[usenames]{xcolor}
\usepackage{graphicx,subcaption}
\usepackage{setspace}
\usepackage{float}
\usepackage{booktabs}
\usepackage{rotating}
\usepackage[vcentermath]{youngtab}

\usepackage[debug,pageanchor=false]{hyperref}
\definecolor{link}{rgb}{.8,.15,.1}
\definecolor{pigment}{rgb}{0.36, 0.54, 0.66}
\definecolor{pigment2}{rgb}{0.19, 0.55, 0.91}
\definecolor{pigment3}{rgb}{0.2, 0.2, 0.6}
\definecolor{light-gray}{gray}{0.75}
\hypersetup{colorlinks=true,linkcolor=link,citecolor=link,urlcolor=link,linktocpage}
\usepackage{cleveref}

\usepackage{array,multirow,booktabs,longtable}
\usepackage{mathrsfs}
\usepackage{tikz-cd} 
\usetikzlibrary{backgrounds, arrows,calc,shapes,decorations.pathreplacing, decorations.markings, automata,positioning,matrix}
\usetikzlibrary{snakes}

\tikzset{
	dot/.style = {circle, fill, minimum size=0.15cm,
		inner sep=0pt, outer sep=0pt},
	O8/.style = {dashed, black, very thick},
	D8/.style = {solid, black, thick},
	D8's/.style = {solid, black, very thick},
	D6/.style = {solid,black,thick}
}

\newcommand\vertarrowbox[3][6ex]{%
  \begin{array}[t]{@{} c c @{}} #2 \\
  \left\downarrow\vcenter{\hrule height #1}\right.\kern-\nulldelimiterspace & #3
  \end{array}%
}

\tikzset{
        cvertex/.style={circle,draw=black,inner sep=1pt,outer sep=3pt},
        vertex/.style={circle,fill=black,inner sep=1pt,outer sep=3pt},
        star/.style={circle,fill=yellow,inner sep=0.75pt,outer sep=0.75pt},
        tvertex/.style={inner sep=1pt,font=\scriptsize},
        gap/.style={inner sep=0.5pt,fill=white}}

\tikzstyle{mybox} = [draw=black, fill=blue!10, very thick,
    rectangle, rounded corners, inner sep=10pt, inner ysep=20pt]
\tikzstyle{boxtitle} =[fill=blue!50, text=white,rectangle,rounded corners]

\def\node#1#2{\overset{#1}{\underset{#2}{\circ}}}
\def\ver#1#2{\overset{{\llap{$\scriptstyle#1$}\displaystyle\circ{\rlap{$\scriptstyle#2$}}}}{\scriptstyle\vert}}

\usetikzlibrary{arrows}
\tikzstyle{every picture}+=[remember picture]
\tikzstyle{na} = [baseline=-.5ex]
\tikzstyle{mine}= [arrows={angle 90}-{angle 90},thick]

\def\Llleftarrow{%
\lower2pt\hbox{\begingroup
\tikz
\draw[shorten >=0pt,shorten <=0pt] (0,3pt) -- ++(-1em,0) (0,1pt) -- ++(-1em-1pt,0) (0,-1pt) -- ++(-1em-1pt,0) (0,-3pt) -- ++(-1em,0) (-1em+1pt,5pt) to[out=-105,in=45] (-1em-2pt,0) to[out=-45,in=105] (-1em+1pt,-5pt);
\endgroup}
}

\newcommand{\cc}{\mathbb{C}}
\newcommand{\zz}{\mathbb{Z}}

\newcommand{\None}{{\mathcal{N}=1}}

\newcommand{\Nfour}{{\mathcal{N}=4}}

\renewcommand{\Re}{\mathrm{Re}}
\renewcommand{\Im}{\mathrm{Im}}

\DeclareMathOperator{\SU}{SU}

\DeclareMathOperator{\U}{U}
\DeclareMathOperator{\SO}{SO}

\DeclareMathOperator{\depth}{depth}

\newcommand{\todo}[1]{}
\renewcommand{\todo}[1]{{\color{red} TODO: {#1}}}
\newcommand{\red}[1]{}
\renewcommand{\red}[1]{{\color{red} {#1}}}
\newcommand{\blue}[1]{}
\renewcommand{\blue}[1]{{\color{blue} {#1}}}

\newcommand{\su}[1]{}
\renewcommand{\su}[1]{{\mathfrak{su}({#1})}}
\newcommand{\uu}[1]{}
\renewcommand{\uu}[1]{{\mathfrak{u}({#1})}}
\newcommand{\so}[1]{}
\renewcommand{\so}[1]{{\mathfrak{so}({#1})}}
\newcommand{\usp}[1]{}
\renewcommand{\usp}[1]{{\mathfrak{usp}({#1})}}

\makeatletter
\renewcommand\xleftrightarrow[2][]{%
  \ext@arrow 9999{\longleftrightarrowfill@}{#1}{#2}}
\newcommand\longleftrightarrowfill@{%
  \arrowfill@\leftarrow\relbar\rightarrow}
\makeatother

\makeatletter
\@addtoreset{equation}{section}
\makeatother

\graphicspath{{figs/}}

\usepackage[en-US,showdow,showzone]{datetime2}
\DTMlangsetup[en-US]{abbr=true,showyear=false,zone=atlantic}

\begin{document}


\begin{titlepage}

\begin{center}

\vskip .3in \noindent


{\Large \textbf{Hierarchies of RG flows in 6d $(1,0)$ massive E-strings}}

\bigskip

{\large \textsc{part ii}}

\bigskip

Marco Fazzi,$^{a}$ Simone Giacomelli,$^{b,c}$ and Suvendu Giri~$^{b,c}$

\bigskip


\bigskip
{\small 

$^a$ Department of Physics and Astronomy,  Uppsala University,  SE-75 120 Uppsala, Sweden \\
\vspace{.25cm}
$^b$ Dipartimento di Fisica, Universit\`a di Milano--Bicocca and \\
$^c$ INFN, sezione di Milano--Bicocca,  \\Piazza della Scienza 3, I-20126 Milano, Italy \\
	
}
\vskip .3cm
{\small \tt \href{mailto:marco.fazzi@physics.uu.se}{marco.fazzi@physics.uu.se} \hspace{.5cm} \href{mailto:simone.giacomelli@unimib.it}{simone.giacomelli@unimib.it} \hspace{.5cm} \href{mailto:suvendu.giri@unimib.it}{suvendu.giri@unimib.it}}

\vskip .6cm
     	{\bf Abstract }
\vskip .1in
\end{center}
\noindent 
We extend the analysis of \href{https://arxiv.org/abs/2208.11703}{arXiv:2208.11703} to the 6d $(1,0)$ SCFTs known as massive E-string theories,  which can be engineered in massive Type IIA with $8-n_0<8$ D8-branes close to an O8$^-$ (or O8$^*$ if $n_0=8,9$).  For each choice of $n_0=1,\ldots,9$ the massive $E_{1+(8-n_0)}$-strings (including the more exotic $\tilde{E}_1$ and $E_0$) are classified by constrained $E_8$ Kac labels, i.e.  a subset of $\text{Hom}(\zz_k,E_8)$,  from which one can read off the flavor subalgebra of $E_{1+(8-n_0)}$ of each SCFT. We construct hierarchies for two types of Higgs branch RG flows: flows between massive theories defined by the same $n_0$ but different labels; flows between massive theories with different $n_0$. These latter flows are triggered by T-brane vev's for the right $\SU$ factor of the SCFT global symmetry, whose rank is a function of both $k$ and $n_0$, a situation which has so far remained vastly unexplored.

\vfill

\begin{flushright}
UUITP-60/22
\end{flushright}
\eject

\end{titlepage}


\tableofcontents
\newpage


\section{Introduction}
\label{sec:intro}

Understanding the structure of the renormalization group (RG) flows that connect conformal field theories is an interesting dynamical question in any dimension. For six-dimensional superconformal field theories (6d SCFTs) the situation is particularly favorable, as these admit multiple string theory engineerings, allowing us to construct and describe these flows rather explicitly.

Six-dimensional SCFTs allow for supersymmetry-preserving deformations triggering an RG flow onto their moduli space \cite{Cordova:2015fha}, whose two main branches are known as tensor branch and Higgs branch.\footnote{There exist also mixed branches, which will not be analyzed here.} The former is parameterized by vacuum expectation values (vev's) for the scalar fields in the tensor multiplets; activating vev's for tensor scalars triggers an RG flow to an infrared (IR) quiver gauge theory (plus tensors) from the ultraviolet (UV) SCFT, which breaks conformality. On the other hand the Higgs branch is parameterized by vev's for the hypermultiplet scalars (i.e. the matter fields), and these trigger flows to a new fixed point in the IR, whose flavor symmetry is generically different from that of the parent UV fixed point, i.e. the symmetry is Higgsed. The Higgsings can be analyzed from a multitude of perspectives:  from nilpotent orbits of Lie algebras (related to Nahm's equations satisfied by the moment maps of the flavor symmetry that rotates the hypermultiplets) \cite{DelZotto:2014hpa,Heckman:2016ssk}, to 3d magnetic quivers \cite{Cabrera:2019izd,Bourget:2019aer}, to the 6d anomaly polynomial \cite{Heckman:2015ola}.

Notable examples of such constructions can be found in \cite{Heckman:2015ola,Heckman:2016ssk,Frey:2018vpw,Giacomelli:2022drw,Fazzi:2022hal}. In particular in \cite{Giacomelli:2022drw,Fazzi:2022hal} the authors of the present paper focused on Higgs branch RG flows for an infinite class of 6d SCFTs known as A-type orbi-instantons. These can be engineered in M-theory, and are the datum of the number $N$ of M5-branes, the order $k$ of the orbifold $\cc^2/\zz_k$ they are probing, and a boundary condition at infinity given by a representation $\rho: \zz_k \to E_8$, where $E_8$ is the gauge symmetry on an M9-wall that the M5's are probing (on top of the orbifold point) \cite{DelZotto:2014hpa}.  For each choice of $(N,k,\rho)$ there exists an orbi-instanton. Then, fixing $k$ one can construct intricate hierarchies of RG flows (either fixing also $N$ as in \cite{Fazzi:2022hal} or letting it change as in \cite{Giacomelli:2022drw}) that connect orbi-instantons defined by different boundary conditions $\rho_i$. The latter are given by a concrete algorithm that takes as input a partition of $k$ in terms of the Coxeter labels $1,\ldots,6,4',3',2'$ of the affine $E_8$ Dynkin diagram (known as a choice of $E_8$ Kac label), and produces as an output a (maximal regular) subalgebra of $E_8$ which is (a factor of) the flavor symmetry algebra of the given orbi-instanton.  For instance for the trivial choice of boundary condition that preserves the full $E_8$ (the associated Kac label is $k=[1^k]$,  adopting standard notation for integer partitions), the full tensor branch of the orbi-instanton is described, in the F-theory language of \cite{Heckman:2015bfa}, by the following electric quiver:
\begin{equation}
[E_8]\ \underbrace{{1}\ \overset{\mathfrak{su}(1)}{2}\ \overset{\mathfrak{su}(2)}{2} \cdots \overset{\mathfrak{su}(k-1)}{2}}_k\ \underbrace{\overset{\su{k}}{\underset{[N_\text{f}=1]}{2}}\ \overset{\su{k}}{2}\cdots \overset{\su{k}}{2}}_{N}\ [\SU(k)]\ .
\end{equation}
The flavor symmetry is thus $E_8 \oplus \su{k}$ (disregarding the $\su{2}$ R-symmetry).\footnote{In this paper we will only deal with the Lie algebra of the flavor symmetry, even when we write the factors as groups.} The right $[\SU(k)]$ is treated as a ``spectator'',  and does not participate in the flows.  (On the contrary, in the present paper this factor will play a crucial role in one type of flows.) Moreover it was conjectured in \cite{Fazzi:2022hal,fazzi-giri-levy} that this hierarchy closely mimics the Hasse diagram of $E_8$-orbits of the double affine Grassmannian of $E_8$ introduced in \cite{Braverman:2007dvq}, which is an appropriate generalization of the $E_8$ affine Grassmannian (based on the \emph{affine} $E_8$ Dynkin rather than the finite one, as for the ``singly'' affine Grassmannian).\footnote{See also \cite{Bourget:2021siw} for another appearance of the affine Grassmannian related to moduli spaces of 3d $\Nfour$ magnetic quivers from brane setups.}

Emboldened by these results, in this paper we will construct a hierarchy of Higgs branch RG flows between 6d SCFTs which are close cousins of the orbi-instantons, but have received relatively less attention in the literature. They are known as massive E-string theories \cite{DelZotto:2014hpa,Zafrir:2015rga,Ohmori:2015tka,Hayashi:2015zka,Bah:2017wxp} (adopting nomenclature introduced in \cite{Bah:2017wxp}),\footnote{Some have been exploited very recently to construct new 4d IR dualities \cite{bajeot-benvenuti}.} since they do not admit an engineering in M-theory but do so both in massive Type IIA --Type IIA in presence of D8-branes sourcing a nonzero Romans mass $F_0$-- and in F-theory (thus falling into the ``atomic classification'' of \cite{Heckman:2015bfa}). Focusing on their massive IIA engineering, these theories are defined by a choice of the number $8-n_0$ of D8's that are on top of an O8$^-$-plane, which is probed by $N$ coincident NS5's intersected by $k$ D6's. (For $n_0=0$ we have 8 D8's on top of the O8$^-$ so that the total Romans mass vanishes and the whole system can be lifted to the original M9-wall of the orbi-instantons \cite{Gorbatov:2001pw}, with the NS5's lifting to M5's and the $k$ D6's to the $\cc^2/\zz_k$ orbifold.) The (at most) $E_8$  flavor symmetry summand of the orbi-instantons is here replaced by $E_{1+(8-n_0)}$,  which for $n_0=1,\ldots,8$ is nothing but the list of exceptional flavor symmetries from \cite{Seiberg:1996bd}. Moreover, if we swap the O8$^-$ with the so-called O8$^*$ \cite{Gorbatov:2001pw}, which has $-9$ D8 charge rather than $-8$ (and is needed to describe the non-perturbative completion of Type I' \cite{Morrison:1996xf,Douglas:1996xp,Cachazo:2000ey}), we are also able to engineer the 6d version of the $\tilde{E}_1=\uu{1}$ (with $n_0=8$) and $E_0=\emptyset$ (with $n_0=9$) theories of \cite{Morrison:1996xf}.

After having introduced the massive E-strings in section \ref{sec:review}, and explained how they are actually related to standard orbi-instantons by a special choice of \emph{constrained} $E_8$ Kac labels, we will construct two types of RG flows connecting them: Higgs branch flows between massive E-strings at fixed $n_0$ but different constrained Kac labels (section \ref{sec:higgsflows}), mimicking closely what done in \cite{Giacomelli:2022drw,Fazzi:2022hal} for orbi-instantons; and Higgs branch flows from massive E-strings at $n_0$ to others at a different $n'_0$ (section \ref{sec:Tflows}), which crucially involve the right $\SU(k)$ factor, i.e. a ``mixing'' between left and right flavor symmetries, which is a novelty with respect to previous literature. These latter flows are triggered by so-called T-brane vev's for matter hypermultiplets charged under $\SU(k)$. We close in section \ref{sec:conc} with some open perspectives. Appendix \ref{app:rules} contains the rules to construct the magnetic quivers of the massive E-strings. 

\section{Orbi-instantons and massive E-string theories}
\label{sec:review}

Let us briefly review the construction of orbi-instantons via Kac labels from \cite{Fazzi:2022hal}. This will allow us to introduce the massive E-string theories as ``fission'' products \cite{Heckman:2018pqx} of orbi-instantons.

An orbi-instanton of type ADE is the 6d $(1,0)$ SCFT that lives on the common worldvolume of $N$ coincident M5-branes probing the intersection between an M9-wall and the orbifold $\cc^2/\Gamma_\text{ADE}$. On top of $N$ and the order of the orbifold, the SCFT is specified by a choice of boundary condition at the spatial infinity $S^3/\Gamma_\text{ADE}$ surrounding the orbifold point, which is a representation $\rho:\Gamma_\text{ADE} \to E_8$.

Let us restrict our attention to type A. For a trivial boundary condition (i.e. the full $E_8$ coming from the M9 is preserved), the partial tensor branch of the SCFT reads:
\begin{equation}\label{eq:Estrnonres}
[E_8]\ \underbrace{\overset{\su{k}}{1}\ \overset{\su{k}}{2} \cdots \overset{\su{k}}{2}}_{N}\ [\SU(k)]\ .
\end{equation}
Upon completely blowing up the base (i.e. going to the generic point on the tensor branch),  we have 
\begin{equation}\label{eq:fullymasslessE8}
[E_8]\ \underbrace{{1}\ \overset{\mathfrak{su}(1)}{2}\ \overset{\mathfrak{su}(2)}{2} \cdots \overset{\mathfrak{su}(k-1)}{2}}_k\ \underbrace{\overset{\su{k}}{\underset{[N_\text{f}=1]}{2}}\ \overset{\su{k}}{2}\cdots \overset{\su{k}}{2}}_{N}\ [\SU(k)]\ .
\end{equation}
As stated, the chosen boundary condition is the one which preserves the full $E_8$ from the left, which can be nicely packaged in a choice of so-called Kac label of $k$, i.e. a partition of the order $k$ of the orbifold in terms of Coxeter labels of the affine $E_8$ Dynkin diagram, with multiplicity (see e.g. \cite[Sec. 2.1]{Fazzi:2022hal}). The generic Kac label thus reads
\begin{equation}\label{eq:longk}
k = [1^{n_1},2^{n_2},3^{n_3},4^{n_4},5^{n_5},6^{n_6}, 4'^{n_{4'}},2'^{n_{2'}},3'^{n_{3'}}]\ ,
\end{equation}
and preserves a (maximal regular) flavor subalgebra $\mathfrak{f}$ of $E_8$ obtained by deleting all nodes in the affine $E_8$ Dynkin diagram
\begin{equation}\label{eq:kaccoord}
\node{}{n_1}-\node{}{n_2}-\node{}{n_3}-\node{}{n_4}-\node{}{n_5}-\node{\ver{}{n_{3'}}}{n_6}-\node{}{n_{4'}}-\node{}{n_{2'}}
\end{equation}
with nonzero multiplicity $n_i$ in \eqref{eq:longk} (including the $n_{i'}$ by abuse of notation), plus potential $\uu{1}$ summands to make the total rank 8.\footnote{See \cite[Sec. 3.3]{Fazzi:2022hal} for the correct physical interpretation of these $\uu{1}$'s, and how to make sense of their ``(de)localization'' along the electric quiver.}  E.g. $k=[1^k]$ preserves the full $E_8$, which is the implicit choice in \eqref{eq:fullymasslessE8}. In fact, there exists \cite[Sec. 7]{Heckman:2015bfa} one distinct F-theory configuration of decorated curves per boundary condition in M-theory.\footnote{Actually, there exist cases where the 6d electric quiver, i.e. the $(1,0)$ quiver gauge theory describing the full tensor branch, can be associated to two distinct SCFTs. The difference manifests itself in the 6d $\theta$ angle of the gauge theory, which can take values $0,\pi$. One SCFT has electric quiver with $\theta=0$, the other the same electric quiver bu $\theta=\pi$. See e.g. \cite[Sec. 4.3]{Fazzi:2022hal}.} (The algorithm to obtain the former was later given in \cite{Mekareeya:2017jgc}. Later \cite{Frey:2018vpw} has explored at length this M/F-theory one-to-one correspondence for all types of orbi-instantons.)

\subsection{Magnetic quivers of orbi-instantons}
\label{sub:magorbi}

Although the SCFTs are strongly coupled, their Higgs branch can be studied in detail via the so-called magnetic quiver, an auxiliary 3d $\Nfour$ quiver gauge theory whose Coulomb branch coincides with the Higgs branch of the 6d theory, not only on the tensor branch but also at its origin, i.e. at ``infinite (gauge) coupling'', where the SCFT resides. 

For orbi-instantons, these magnetic quivers have been worked out in \cite{Cabrera:2019izd}, and can be read off of the spectrum of D1-branes suspended between a system of NS5-D5 branes, obtained by applying S-duality (i.e. mirror symmetry) to the triple T-duality of the O8-D8-D6-NS5 Type IIA reduction of the M-theory setup. They are given by star-shaped quivers of unitary gauge groups with three tails, obtained by gluing an affine $E_8$ Dynkin diagram to the $T(\SU(k))$ 3d theory of \cite{Gaiotto:2008ak}:
\begin{equation}\label{eq:genericmagquiv}
 {
 1 - 2 - \cdots - (k-1) - k - r_1 -r_2 -r_3 - r_4 -r_5-\overset{\overset{\displaystyle r_{3'}}{\vert}}{r_6}-r_{4'}-r_{2'}}\ ,
\end{equation}
where $-$ or $|$ denotes a hypermultiplet,  $n$ a $\U(n)$ gauge group, and the ranks $r_i, r_{i'}$ depend nontrivially on $N$ and the choice of Kac label. The gluing is done ``along'' the extending node of affine $E_8$.

\subsection{\texorpdfstring{Massive E-string theories for $E_8,\ldots,E_1$}{Massive E-string theories for E8,...,E1}}
\label{sub:massive}

\begin{figure}[th!]
\centering

\begin{tikzpicture}[scale=1,baseline]
\node at (0,0) {};
\draw[fill=black] (0.5,0) circle (0.075cm);
\draw[fill=black] (1.5,0) circle (0.075cm);
\draw[fill=black] (2.5,0) circle (0.075cm);
\draw[fill=black] (3.5,0) circle (0.075cm);
\draw[fill=black] (4.5,0) circle (0.075cm);
\draw[fill=black] (5.5,0) circle (0.075cm);
\draw[fill=black] (6.5,0) circle (0.075cm);
\draw[fill=black] (7.5,0) circle (0.075cm);
\draw[fill=black] (8.5,0) circle (0.075cm);


\draw[solid,black,thick] (0.5,0)--(1.5,0) node[black,midway,yshift=0.2cm] {\footnotesize $1$};
\draw[solid,black,thick] (1.5,0)--(2.5,0) node[black,midway,yshift=0.2cm] {\footnotesize $2$};
\draw[loosely dotted,thick,black] (2.5,0)--(3.5,0) node[black,midway] {};
\draw[solid,black,thick] (3.5,0)--(4.5,0) node[black,midway,yshift=0.2cm] {\footnotesize $k-1$};
\draw[solid,black,thick] (4.5,0)--(5.5,0) node[black,midway,xshift=0.2cm,yshift=0.2cm] {\footnotesize $k$};
\draw[solid,thick,black] (5.5,0)--(6.5,0) node[black,midway,yshift=0.2cm] {\footnotesize $k$};
\draw[loosely dotted,thick,black] (6.5,0)--(7.5,0) node[black,midway] {};
\draw[solid,black,thick] (7.5,0)--(8.5,0) node[black,midway,yshift=0.2cm] {\footnotesize $k$};
\draw[solid,black,thick] (8.5,0)--(9.5,0) node[black,midway,yshift=0.2cm] {\footnotesize $k$};

\draw[dashed,black,very thick] (0,-.5)--(0,.5) node[black,midway, xshift =0cm, yshift=-1.5cm] {} node[black,midway, xshift =0cm, yshift=1.5cm] {};
\draw[solid,black,very thick] (0.25,-.5)--(0.25,.5) node[black,midway, xshift =0cm, yshift=+.75cm] {\footnotesize $7$};
\draw[solid,black,very thick] (4.5+0.5,-.5)--(4.5+0.5,.5) node[black,midway, xshift =0cm, yshift=+.75cm] {\footnotesize $1$};
\end{tikzpicture}
\caption{Type IIA configuration dual to the F-theory one in \eqref{eq:fullymasslessE8}. The SCFT corresponds to the configuration where all NS5's are brought on top of each other. Circles represent NS5's; vertical lines represent D8's, with their total number written on top; a dashed vertical line represents the O8$^-$; horizontal lines represent D6's, with their total number written on top.}
\label{fig:E8-IIA}
\end{figure}
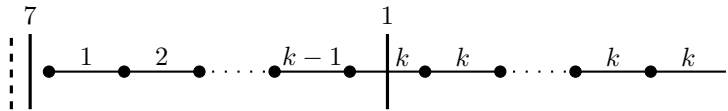

The F-theory configuration in \eqref{eq:fullymasslessE8} allows for an alternative engineering in Type IIA, given by stacking 7 D8-branes on top of an O8$^-$ on the left, plus a lone D8 far away from it (corresponding to the single fundamental hypermultiplet at the beginning of the plateau in position $k$). See figure \ref{fig:E8-IIA}. In a sense, we are stripping off $n_0=1$ D8 from a stack of 8 which, together with the O8$^-$, lifts to the M9-wall at strong string coupling (see the discussion in \cite[Sec. 2.2]{Fazzi:2022hal}). 

What happens if we instead strip off $n_0>1$ D8's, and clump them together? We land on the electric quiver \cite[Eq. (5.71)]{Mekareeya:2017jgc}
\begin{equation}\label{eq:massEntail}
[E_{1+(8-{n}_0)}]\ 1\ \overset{\mathfrak{su}({n}_0)}{2}\ \overset{\mathfrak{su}(2{n}_0)}{2}\ \overset{\mathfrak{su}(3{n}_0)}{2} \cdots \ \overset{\mathfrak{su}((m-1){n}_0)}{2} \ \underset{[N_\text{f}={n}_0]}{\overset{\mathfrak{su}(m{n}_0)}{2}}\ \overset{\mathfrak{su}(m{n}_0)}{2} \cdots \overset{\mathfrak{su}(m{n}_0)}{2} \ [\SU(m{n}_0)]\ ,
\end{equation}
where now $k=mn_0$, and $E_{1+(8-{n}_0)}$ on the left is the following list of flavor algebras:
\begin{equation}\label{eq:seiberglist}
E_{1+(8-n_0)} = \{ E_8, E_7, E_6, \mathfrak{so}(10),\mathfrak{su}(5),\mathfrak{su}(3)\oplus \mathfrak{su}(2), \mathfrak{su}(2)\oplus \mathfrak{u}(1), \mathfrak{su}(2)\}_{n_0=1}^{8}\ .
\end{equation}
Notice that in the above electric quiver we still have a total of $8-n_0+n_0=8$ D8's, i.e. the total Romans mass of the system vanishes (since the O8$^-$ contributes $-8$).  Moreover $N$ and $k$ are required to satisfy the ``massless constraint'', namely that the total number of compact curves $12\ldots 2$ be larger than $k$.  (In \cite{Fazzi:2022hal} that number was called $N$.)

The list in \eqref{eq:seiberglist} is motivated by a similar analysis in 5d \cite{Seiberg:1996bd}, where probe D4's are brought close to an O8$^-$ with $8-n_0$ D8's on top.\footnote{$E_{1+(8-n_0)}$ is the enhanced exceptional flavor symmetry of a 5d SCFT which flows in the IR to 5d $\None$ $\SU(2)$ SYM with $N_\text{f} = 8-n_0$ fundamentals.} This excludes the $\tilde{E}_1 = \uu{1}$ and $E_0 = \emptyset$ cases later discovered in \cite{Morrison:1996xf}, which we will comment on momentarily. Following the algorithm in \cite{Mekareeya:2017jgc} we recognize that \eqref{eq:massEntail} is the right electric quiver only for $n_0=1,\ldots,6$ with Kac label $k=mn_0=[n_0^m]$ and preserved flavor algebra $\mathfrak{f}$ given by, respectively:
\begin{equation}\label{eq:massEntailcorr}
n_0=1,\ldots,6: \mathfrak{f}= E_{1+(8-n_0)} \oplus \begin{cases} \text{``$\su{1}$''}\\ \su{2} \\ \su{3} \\ \su{4} \\ \su{5} \\ \su{6}  \end{cases}\hspace{-.5cm} : \ [E_{1+(8-{n}_0)}]\ 1\ \overset{\mathfrak{su}({n}_0)}{2} \cdots \underset{[N_\text{f}={n}_0]}{\overset{\mathfrak{su}(m{n}_0)}{2}}\ \cdots [\SU(m{n}_0)]\ .
\end{equation}
Other Kac labels involving the primed Coxeter labels $2',3',4'$ exist of course, and have a preserved flavor algebra $\mathfrak{f}_\text{primed}$ which contains a \emph{subalgebra} of the $E_{1+(8-{n}_0)}$ summand in the $\mathfrak{f}_\text{unprimed}=E_{1+(8-{n}_0)}\oplus \su{n_0}$ of \eqref{eq:massEntailcorr}.  For this reason they will lie lower on the hierarchy of RG flows between massive strings, i.e. they are never the starting points of the flows, as we will see in section \ref{sec:higgsflows}.

For $n_0=7$ we can have multiple electric quivers (or labels) which preserve the same $\mathfrak{f}=E_2 \oplus \su{7} = \su{2}\oplus \uu{1}\oplus \su{7}$,  depending on the value of $k$ (which is \emph{not} necessarily given by $7m$ now). Adapting from \cite[Sec. 4.1]{Fazzi:2022hal} we obtain:
\begin{subequations}
\label{eq:E2quivs}
\begin{align} 
&k=[4^{n_{4'}},3^{n_{3'}},2^{n_{2'}}]:\ [E_2]\ \overset{\usp{2n_{2'}}}{1}\ \overset{\su{2n_{2'}+7}}{2}\ \cdots\ \overset{\su{k}}{\underset{[N_\text{f}=7]}{2}} \ [\SU(k)]\ , \\
&k=[4^{n_{3'}-1},3^{n_{3'}},2^{n_{2'}}]:\ [E_2]\ \overset{\su{2n_{2'}+3}}{\underset{[N_{\fontsize{0.5pt}{1pt}\selectfont \yng(1,1)}=1]}{1}}\ \overset{\su{2n_{2'}+3+7}}{2}\ \cdots\ \overset{\su{k}}{\underset{[N_\text{f}=7]}{2}} \ [\SU(k)]\ , \\
&k=[4^{n_{3'}+1},3^{n_{3'}},2^{n_{2'}}]:\ [E_2]\ \overset{\su{2n_{2'}+4}}{\underset{[N_{\fontsize{0.5pt}{1pt}\selectfont \yng(1,1)}=1]}{1}}\ \overset{\su{2n_{2'}+4+7}}{2}\ \cdots\ \overset{\su{k}}{\underset{[N_\text{f}=7]}{2}} \ [\SU(k)]\ .
\end{align}
\end{subequations}
Here and below $N_{\fontsize{0.25pt}{0.5pt}\selectfont \yng(1,1)}=1$ indicates a single two-index antisymmetric hypermultiplet of $\su{r \geq 3}$, whereas $N_{\fontsize{0.25pt}{0.5pt}\selectfont \tfrac{1}{2} \yng(1,1,1)}=1$ a half-hypermultiplet in the three-index antisymmetric of $\su{6}$. 

For $n_0=8$ we have $\mathfrak{f}= E_1 \oplus \su{8} = \su{2}\oplus \su{8}$ and
\begin{subequations}
\label{eq:E1quivs}
\begin{align} 
&k=[4^{n_{4'}},2^{n_{2'}}]\ \text{with $n_{4'}$ even}:\ [E_1]\ \overset{\usp{2n_{2'}}}{1}\ \overset{\su{2n_{2'}+8}}{2}\ \cdots\ \overset{\su{k}}{\underset{[N_\text{f}=8]}{2}} \ [\SU(k)]\ , \\
&k=[4^{n_{4'}},2^{n_{2'}}]\ \text{with $n_{4'}$ odd}:\ [E_1]\ \overset{\su{2n_{2'}+4}}{\underset{[N_{\fontsize{0.5pt}{1pt}\selectfont \yng(1,1)}=1]}{1}}\ \overset{\su{2n_{2'}+12}}{2}\ \cdots\ \overset{\su{k}}{\underset{[N_\text{f}=8]}{2}} \ [\SU(k)]\ , 
\end{align}
\begin{align}
&k=[3^{n_{3'}},2^{n_{2'}}]\ \text{with} \begin{cases}  n_{3'}\ \text{even} \\ n_{2'} \geq \tfrac{n_{3'}}{2} \end{cases} \hspace{-.5cm}: [E_1]\ \overset{\usp{2n_{2'}-n_{3'}}}{1}\ \overset{\su{2n_{2'}-n_{3'}+8}}{2}\ \cdots\ \overset{\su{k}}{\underset{[N_\text{f}=8]}{2}} \ [\SU(k)]\ , \\
&k=[3^{n_{3'}},2^{n_{2'}}]\ \text{with} \begin{cases}  n_{3'}\ \text{odd} \\ n_{2'} \geq \tfrac{n_{3'}-1}{2} \end{cases}\hspace{-.5cm}:\ [E_1]\ \overset{\su{2n_{2'}-n_{3'}+4}}{\underset{[N_{\fontsize{0.5pt}{1pt}\selectfont \yng(1,1)}=1]}{1}}\ \overset{\su{2n_{2'}-n_{3'}+8}}{2}\ \cdots\ \overset{\su{k}}{\underset{[N_\text{f}=8]}{2}} \ [\SU(k)]\ .
\end{align}
\end{subequations}
After much work (albeit straightforward), we are finally ready to introduce the massive E-string theories, which as the name suggests do not have an engineering in M-theory but do so in massive Type IIA (and F-theory). In fact they are generalizations of \eqref{eq:massEntailcorr}, \eqref{eq:E2quivs}, \eqref{eq:E1quivs} obtained by ``cutting'' the electric quiver right at the beginning of the plateau (i.e. where the $N_\text{f}=n_0$ fundamental hypermultiplets are located).  All cases with empty $-1$ curve (which is the only possibility for $n_0=1,\ldots,6$ but is only a limiting case of one electric quiver for $n_0=7,8$) have already appeared in \cite[Sec. 5.2]{Bah:2017wxp}. Here we will extend the notion of massive theory to all electric quivers.

The cutting operation leaves behind only $8-n_0$ D8's (rather than 8), and the total Romans mass is no more vanishing. We can think of this ``fission'' \cite{Heckman:2018pqx} as the ungauging of the first $\su{k}$ algebra (counting from the left), obtained in the IIA picture by semi-infinitely stretching the $k$ D6's in that finite segment (see again figure \ref{fig:E8-IIA}) to the far right (thus pushing away to infinity the $N_\text{f}=n_0$ D8's that used to cross them),  a picture that will become useful in what follows. (Equivalently, in the F-theory picture we are decompactifying the $-2$ curve that supports $\su{k}$,  becoming $[\SU(k)]$ in our notation.)

E.g. for $n_0=1,\ldots,6$ we are left with the shortened electric quiver \cite[Eq. (5.11)]{Bah:2017wxp}
\begin{equation}\label{eq:massive}
[E_{1+(8-n_0)}]\ \underbrace{{1}\ \overset{\mathfrak{su}(n_0)}{2}\ \overset{\mathfrak{su}(2n_0)}{2} \cdots \overset{\mathfrak{su}((N-1)n_0)}{2}}_{N}\  [\SU(Nn_0)]\ ,
\end{equation}
upon renaming $m=N$ and $k=Nn_0$.  (We will sometimes refer to the above theory as massive $E_{1+(8-n_0)}$-string or massive string at $n_0$, if we want to stress the choice of $n_0$. Remember that for $n_0=7,8$ we have multiple electric quivers depending on the value of $k$ for the parent orbi-instanton.) This means the massless constraint is no longer satisfied in the massive theories; rather, $k$ (which has lost its meaning as order of the orbifold) is given in terms of $N$ and signifies the number of semi-infinite D6's in the last segment to the right.  $N$ is then the total number of compact curves $1 2\ldots2$ (as in the notation of \cite{Fazzi:2022hal}).

\subsection{\texorpdfstring{The O8$^*$-plane and the 6d massive $\tilde{E}_1,E_0$ theories}{The O8*-plane and the 6d massive E1,E0 theories}}
\label{sub:O8*}

Studying the 5d SCFTs which come from D4 probes of an O8-D8 system in Type I', \cite{Morrison:1996xf} discovered they could add two extra cases, $\tilde{E}_1 = \uu{1}$ and $E_0 = \emptyset$, to the list in \eqref{eq:seiberglist} if they considered a type of orientifold 8-plane dubbed O8$^*$\footnote{This notation was introduced later in \cite{Gorbatov:2001pw}. This orientifold also plays a role in getting the $\SO(34)$ and $\SU(18)$ gauge groups in Type I' \cite{Bachas:1997kn}.} which is necessary to understand the non\hyp{}perturbative completion of Type I' \cite{Morrison:1996xf,Douglas:1996xp,Cachazo:2000ey}. The O8$^*$ has $-9$ D8 charge, and can be thought of as the product of nucleating an extra D8 out of an O8$^-$ (which has charge $-8$), that is (schematically) $\text{O8}^- \to\ \text{O8}^* + 1\ \text{D8}$ (which reminds of the well-established Type IIB splitting $\text{O7}^- \to [1,1]+[1,-1]$ 7-branes \cite{sen-O7}).  

We propose the following picture in massive IIA, which extends naturally the definition of \cite{Bah:2017wxp} (only valid down to $E_1$):
\begin{itemize}
\item $E_1$ is given by an O8$^-$ with $8-n_0=0$ D8's on  top (i.e.  we stripped off $n_0=8$ D8's from it, and moved them inside the massless electric quiver \eqref{eq:massEntail}, in position $m$);
\item $\tilde{E}_1$ is given by an O8$^*$ with $1$ extra D8, the one nucleated by the O8$^-$, i.e.  $8-n_0=0$ or $n_0=8$. This system obviously has the same charge as an O8$^-$ with $0$ D8's,  i.e. case $E_1$ above. However there is a nonzero $\theta$ angle in massive IIA that distinguishes the two (see below);
\item $E_0$ is given by the O8$^*$ alone, i.e.  $8-n_0=-1$ or $n_0=9$, with nonzero $\theta$ angle.
\end{itemize}
To explain our reasoning, we take a short historical detour.\footnote{We would like to thank Oren Bergman for suggesting some of the arguments below and especially for sharing with us the unpublished notes for \cite{orentalk}.}

It has been proposed in \cite{Sethi:2013hra} that the 10d Type I string, i.e. the orientifold (O9) of Type IIB (with 32 D9's),  comes in two flavors: the ordinary string with $C_0=0$ and a ``new'' one with $C_0=\tfrac{1}{2}$. This is because the orientifold imposes $C_0 \to -C_0$, but $C_0$ is periodic in IIB (it is an axion), $C_0 \sim C_0 +1$,\footnote{This is imposed by the $\text{SL}(2,\zz)$ gauge symmetry of IIB, acting on $\tau=C_0+\frac{i}{g_\text{s}}$ as $\tau \to \frac{a\tau +b}{c\tau +d}$.  From this perspective $C_0 \to C_0+1$ is nothing but a $T = \left( \begin{smallmatrix} 1 & 1 \\ 0 & 1 \end{smallmatrix} \right)$ transformation.} therefore both $C_0=0,\tfrac{1}{2}$ are allowed.  $C_0$ is then interpreted as a $\zz_2$-valued $\theta$ angle in 10d. This possibility was explored long ago in \cite[footnote 8]{Bergman:2000tm} (where the two 10d backgrounds are shown to be connected by a non-BPS D8-brane coming from a D9-$\overline{\text{D9}}$ pair \cite{Bergman:1999ta}), though the connection to $C_0$ was made only in \cite{Sethi:2013hra}. The correct interpretation of the new string was then given in \cite{orentalk}, which showed the two are actually completely equivalent.  (The conclusion of \cite{orentalk} is that the orientifold of Type IIB with $C_0=\tfrac{1}{2}$ is Type I with the other chirality of the $\text{Spin}(32)$ spinor.) Much more recently \cite{Montero:2022vva} arrived at the same conclusion based on arguments due to \cite{Witten:1998cd}. (The argument goes roughly as follows. The gauge group of Type I is actually $\text{Spin}(32)/\zz_2$ due to D$(-1)$-instantons. There are two possible $\zz_2$ projections which are exchanged by an element in the disconnected component of O$(32)$. However the two choices of gauge symmetry are equivalent, since any sign weighting the instanton contribution in the 10d action is identified with the other by the same element.)

In 9d Type I', i.e. the orientifold (O8) of Type IIA on a circle (with 16 D8's),  which is T-dual to Type I, the orientifold forces $\int_{S^1} C_1=0,\tfrac{1}{2}$,  thus $\int C_1$ plays the role of a 9d $\zz_2$-valued $\theta$ angle \cite{Bergman:2013ala}. (Moreover this 9d $\theta$ is naturally interpreted as the one distinguishing 5d $\SU(2)_0$ and $\SU(2)_\pi$ pure SYM, to which the $E_1$ and $\tilde{E}_1$ 5d SCFTs flow, respectively.) Here we have two O8$^-$-planes at the ends of the interval $S^1/\zz_2$, plus 16 D8's ensuring D8 charge cancellation in this compact space. The domain wall that separates the two backgrounds (at different $\theta$) is a non-BPS D7-brane coming from a D8-$\overline{\text{D8}}$ pair. If we instead add in an extra D8 (i.e. an extra flavor in 5d) and bring it close to the O8, the system is unstable and the non-BPS D7 (that used to be on top of the O8) is absorbed into the D8, destroying the domain wall (see \cite[Fig. 3]{Bergman:2013ala}). In other words, extra D8's make the 9d $\theta$ angle unphysical (as the 10d $\theta$ in Type I).  On the other hand without extra D8's the angle is physical.

From this perspective, the 5d $E_1$ theory has zero D8's on the (say) left O8.  Massless D0's are responsible for the enhancement of the global symmetry in the 5d D4 probe theory to $\su{2}$. Taking now one of the D8's from the other stack of 16 all the way around the circle (i.e. across the left O8 and back to the right one) creates a fundamental string between a stuck D0 and the O8,\footnote{D0's in a massive background necessarily have strings attached. See e.g. \cite[Sec. 3.3.2]{Bergman:2020bvi}.} so that the former is no more massless, and there is no enhancement of global symmetry, only a $\uu{1}$. That is, we landed on $\tilde{E}_1$.  

Recently \cite{Montero:2022vva} has linked the enhancement to the vanishing of the 9d $\theta$ angle $\int_{S^1} C_1$:
\begin{itemize}
\item when $\int_{S^1} C_1=0$, the O8$^-$ has massless D0-branes on top, and the flavor symmetry algebra (engineered by  the $n_0=1,\ldots,8$ D8's which are still on the O8) enhances to an exceptional one \cite{Polchinski:1995df,Matalliotakis:1997qe,Bergman:1997py} as in \eqref{eq:seiberglist}.\footnote{See the explanation e.g. on \cite[p.9]{Fazzi:2022hal}.}
\item When $\int_{S^1} C_1=\frac{1}{2}$ instead, the O8$^-$ cannot host massless D0's,  but can emit non\hyp{}perturbatively an extra D8, i.e. becomes the O8$^*$, giving rise to the $\tilde{E}_1$ ($n_0=8$) and $E_0$ ($n_0=9$) cases.
\end{itemize}
Our setup is in massive IIA rather than I'. This is equivalent to stretching the interval of I' to infinite length, or simply to disregarding one of the two ends and some out of the 16 D8's.  It is possible that an analog 10d $\theta$ angle exists in massive IIA,  and selects the O8$^*$ when it attains one of its two possible values. We will \emph{assume} this is the case, since the existence of orbi-instantons constructed with the O8$^*$ naturally suggests the existence of massive E-strings with the same ingredient via fission.  

\begin{figure}[ht!]
\centering
\begin{tikzpicture}[scale=1,baseline]
				\draw[fill=black] (0,0) circle (0.075cm);
				\draw[fill=black] (1,0) circle (0.075cm);
				\draw[fill=black] (2,0) circle (0.075cm);
				\draw[fill=black] (3,0) circle (0.075cm);
				\draw[fill=black] (4.5,0) circle (0.075cm);
				
				\draw[solid,black,thick] (1,0)--(2,0) node[black,midway,yshift=0.2cm] {\footnotesize $9$};
				\draw[loosely dotted,black,thick] (2,0)--(3,0) node[black,midway,yshift=0.2cm] {};
				\draw[solid,black,thick] (3,0)--(4.5,0) node[black,midway,yshift=0.2cm] {\footnotesize $k-9$};
				\draw[solid,black,thick] (4.5,0)--(5.5,0) node[black,midway,xshift=0cm,yshift=0.2cm] {\footnotesize $k$};
				
				\draw[dotted,black,very thick] (0,-.5)--(0,.5) node[black,midway,yshift=+.75cm] {\footnotesize O8$^*$};
\end{tikzpicture}
\hspace*{2cm}
\begin{tikzpicture}[scale=1,baseline]
				\draw[fill=black] (0,0) circle (0.075cm) node[xshift=-0.3cm] {${\fontsize{0.25pt}{0.5pt}\selectfont \yng(1,1)}$};
				\draw[fill=black] (1,0) circle (0.075cm);
				\draw[fill=black] (2,0) circle (0.075cm);
				\draw[fill=black] (3,0) circle (0.075cm);
				\draw[fill=black] (4.5,0) circle (0.075cm);
				
				\draw[solid,black,thick] (0,0)--(1,0) node[black,midway,yshift=0.2cm] {\footnotesize $3$};
				\draw[solid,black,thick] (1,0)--(2,0) node[black,midway,yshift=0.2cm] {\footnotesize $12$};
				\draw[loosely dotted,black,thick] (2,0)--(3,0) node[black,midway,yshift=0.2cm] {};
				\draw[solid,black,thick] (3,0)--(4.5,0) node[black,midway,yshift=0.2cm] {\footnotesize $k-9$};
				\draw[solid,black,thick] (4.5,0)--(5.5,0) node[black,midway,xshift=0cm,yshift=0.2cm] {\footnotesize $k$};
				
				\draw[dotted,black,very thick] (0,-.5)--(0,.5) node[black,midway,yshift=+.75cm] {\footnotesize O8$^*$};
\end{tikzpicture}
\vspace{.5cm}
\begin{tikzpicture}[scale=1,baseline]
				\draw[fill=black] (0,0) circle (0.075cm) node[xshift=-0.4cm] {$\tfrac{1}{2} {\fontsize{0.25pt}{0.5pt}\selectfont \yng(1,1,1)}$};
				\draw[fill=black] (1,0) circle (0.075cm);
				\draw[fill=black] (2,0) circle (0.075cm);
				\draw[fill=black] (3,0) circle (0.075cm);
				\draw[fill=black] (4.5,0) circle (0.075cm);
				
				\draw[solid,black,thick] (0,0)--(1,0) node[black,midway,yshift=0.2cm] {\footnotesize $6$};
				\draw[solid,black,thick] (1,0)--(2,0) node[black,midway,yshift=0.2cm] {\footnotesize $15$};
				\draw[loosely dotted,black,thick] (2,0)--(3,0) node[black,midway,yshift=0.2cm] {};
				\draw[solid,black,thick] (3,0)--(4.5,0) node[black,midway,yshift=0.2cm] {\footnotesize $k-9$};
				\draw[solid,black,thick] (4.5,0)--(5.5,0) node[black,midway,xshift=0cm,yshift=0.2cm] {\footnotesize $k$};
				
				\draw[dotted,black,very thick] (0,-.5)--(0,.5) node[black,midway,yshift=+.75cm] {\footnotesize O8$^*$};
\end{tikzpicture}
\caption{Type IIA configurations with O8$^*$ engineering the $E_0$ theory with $n_0=9$ and $k=[3^{n_{3'}}]$. \textbf{Top left:} $n_{3'}=0\mod 3$. \textbf{Top right:} $n_{3'}=1\mod 3$. \textbf{Bottom:} $n_{3'}=2\mod 3$.}
\label{fig:IIAE0}
\end{figure}
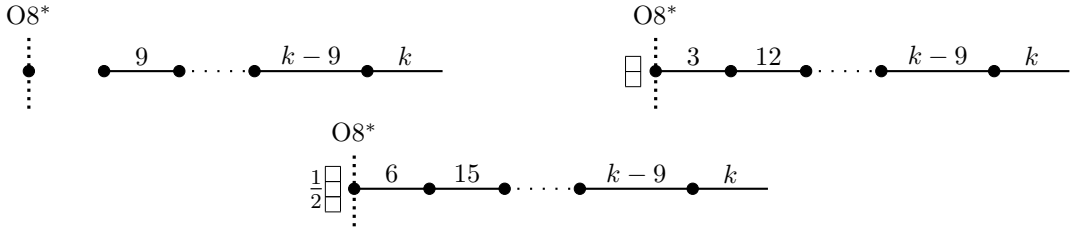

Following once again the algorithm in \cite{Mekareeya:2017jgc} and adapting from \cite[Sec. 4.1]{Fazzi:2022hal},  for $n_0=8$ we have $\mathfrak{f}=\tilde{E}_1 \oplus \su{8}=\uu{1} \oplus \su{8}$ and
\begin{subequations}
\label{eq:tildeE1quivs}
\begin{align} 
&k=[3^{2n_{2'}},2^{n_{2'}}]:\ [\tilde{E}_1]\ 1 \ \overset{\su{8}}{2}\ \overset{\su{16}}{2}\ \cdots\ \overset{\su{k}}{\underset{[N_\text{f}=8]}{2}} \ [\SU(k)]\ , \\
&k=[3^{2n_{2'}+1},2^{n_{2'}}]:\ [\tilde{E}_1]\ \overset{\su{3}}{\underset{[N_{\fontsize{0.5pt}{1pt}\selectfont \yng(1,1)}=1]}{1}}\ \overset{\su{11}}{2}\  \overset{\su{19}}{2}\ \cdots\ \overset{\su{k}}{\underset{[N_\text{f}=8]}{2}} \ [\SU(k)]\ , \\
&k=[3^{2n_{2'}+2},2^{n_{2'}}]:\ [\tilde{E}_1]\ \overset{\su{6}}{\underset{[N_{\fontsize{0.5pt}{1pt}\selectfont \tfrac{1}{2} \yng(1,1,1)}=1]}{1}}\ \overset{\su{14}}{2}\  \overset{\su{22}}{2}\ \cdots\ \overset{\su{k}}{\underset{[N_\text{f}=8]}{2}} \ [\SU(k)]\ ,
\end{align}
\end{subequations}
while for $n_0=9$ we have $\mathfrak{f}=E_0 \oplus \su{9} = \su{9}$ and
\begin{subequations}
\label{eq:E0quivs}
\begin{align} 
&k=[3^{2n_{3'}}]\ \text{with $n_{3'}=0 \mod 3$}:\ [E_0]\ 1 \ \overset{\su{9}}{2}\ \overset{\su{18}}{2}\ \cdots\ \overset{\su{k}}{\underset{[N_\text{f}=9]}{2}} \ [\SU(k)]\ , \\
&k=[3^{2n_{3'}}]\ \text{with $n_{3'}=1 \mod 3$}:\ [E_0]\ \overset{\su{3}}{\underset{[N_{\fontsize{0.5pt}{1pt}\selectfont \yng(1,1)}=1]}{1}}\ \overset{\su{12}}{2}\  \overset{\su{21}}{2}\ \cdots\ \overset{\su{k}}{\underset{[N_\text{f}=9]}{2}} \ [\SU(k)]\ , \\
&k=[3^{2n_{3'}}]\ \text{with $n_{3'}=2 \mod 3$}:\ [E_0]\ \overset{\su{6}}{\underset{[N_{\fontsize{0.5pt}{1pt}\selectfont \tfrac{1}{2} \yng(1,1,1)}=1]}{1}}\ \overset{\su{15}}{2}\  \overset{\su{24}}{2}\ \cdots\ \overset{\su{k}}{\underset{[N_\text{f}=9]}{2}} \ [\SU(k)]\ .
\end{align}
\end{subequations}
Cutting at the beginning of the plateau (where the $N_\text{f}=n_0$ fundamentals are located) we obtain the corresponding E-strings.  See figure \ref{fig:IIAE0} for a brane realization of the $E_0$ case for instance.

\subsubsection{5d SCFTs from massive E-strings}
\label{subsub:5d}

We close this section with an observation which begs for further investigation.  It is interesting to ask whether the 6d massive E-strings give rise upon compactification to new (high-rank) 5d SCFTs according to the arguments in \cite{DelZotto:2017pti,Jefferson:2017ahm,Jefferson:2018irk}.\footnote{E.g. the \emph{massless} rank-1 E-string gives rise upon compactification to the Seiberg list of 5d SCFTs with exceptional flavor symmetry given by $E_{1+(8-n_0)},\tilde{E}_1,E_0$; see e.g. the left column in \cite[Fig. 10]{Jefferson:2018irk}. For orbi-instantons, i.e.  rank-$N$ E-strings with decorated base curves, this question has received a partial answer in \cite[Sec. 5.3.2]{DelZotto:2017pti}. The authors expect they will indeed give rise to 5d SCFTs.  One possible way to approach the problem is to use the combined fiber diagram (CFD) technology of \cite{Apruzzi:2018nre,Apruzzi:2019vpe,Apruzzi:2019opn,Apruzzi:2019enx} by gluing the CFD for $(E_8,A)$ conformal matter provided in \cite[Sec. 4.10 \& 4.11]{Apruzzi:2019enx} to that for the $(A,A)$ bifundamentals in \cite[Tab. 5]{Apruzzi:2019kgb}, since these are the two building blocks of any orbi-instanton. The same CFD technology could also prove useful in the massive case. (Other ways of getting 5d SCFTs from 6d include \cite{Bhardwaj:2020ruf,Bhardwaj:2020avz,Bhardwaj:2019jtr}.)} One may also be tempted to consider more than 8 D8's on the O8$^-$ or even the O8$^+$ to construct even more massive 6d SCFTs, given the Romans mass can be progressively increased with these two ingredients.  (That both possibilities give rise to 6d SCFTs can be checked holographically \cite{Apruzzi:2017nck}.) The flavor symmetry algebra is then of type D (O8$^-$ + $n$ D8's) and C (O8$^+$ + $n$ D8's) respectively, rather than E.  (Type A corresponds to the case with only D8's, and is the well-studied NS5-D6-D8 setup of \cite{Hanany:1997gh,Brunner:1997gk}.) However the arguments of \cite{Seiberg:1996bd} show that considering those possibilities leads to UV \emph{free} theories rather than CFTs in 5d (see e.g. \cite[Tab. 1]{Bedroya:2021fbu}). One last possibility is represented by the O8$^0$ and O8$^{-1}$ planes of \cite{Keurentjes:2000bs,Aharony:2007du}, of D8 charge $0$ and $-1$ respectively,  the latter being the non\hyp{}perturbative completion of the shift orientifold O8$^0$ in Type I' (i.e.  O8$^0 \to\ \text{O8}^{-1} + 1\ \text{D8}$). Here the swampland arguments of \cite{Bedroya:2021fbu} suggest the O8$^{-1}$ should give rise to a 5d SCFT with empty flavor symmetry (as in the $E_0$ case) which can be reached via a relevant deformation from the $\SU(2)_\pi$ gauge theory with an adjoint.  In turn this latter KK theory is the twisted compactification of the $A_1$ $(2,0)$ \cite[Eq. (B.3)]{Bhardwaj:2019fzv}, i.e. is UV-completed in 6d.\footnote{At infinite $g_\text{s}^\text{I'}$ coupling, the O8$^0$ emits non\hyp{}perturbatively a D8,  i.e. splits as O8$^{-1}$ $+$ 1 D8. This system lifts to M-theory on a Klein bottle, which can be thought of (in the appropriate region in moduli space) as a long tube with two cross-caps at the ends \cite{Aharony:2007du}. It is reasonable then that a single M5, whose worldvolume theory is the $A_1$ $(2,0)$,  yields upon twisted compactification on this long tube said KK theory. Given the latter has a 5d $\theta=\pi$, and that this is identified with the 9d $\theta$ of Type I' \cite[Eq. (2)]{Bergman:2013ala}, we learn that a background with the O8$^{-1}$ has nonzero $\theta$.} When dealing with honest Type I' setups, there are limitations to the combinations of 8-planes one can place at the ends of the interval $S^1/\zz_2$. However in massive IIA we can consider only one of the two ends, and ask whether combinations of 8-planes and branes placed at the origin of the semi-infinite line give rise to new interesting 6d SCFTs (which at any rate must admit an F-theory engineering contemplated in \cite{Heckman:2015bfa}).

\subsection{Constrained $E_8$ Kac labels}
\label{sub:constrainedkac}

At this point one may wonder what is left of the Kac labels used to classify different orbi-instantons (and their magnetic quivers), whose distinct electric quivers\footnote{Modulo 6d $\theta$ angle.} give rise to \eqref{eq:massive} and analogs with nonempty $-1$ curve,  upon fission. After all, it seems that the electric quiver \eqref{eq:fullymasslessE8} for $[1^k]$, which we know preserves the full $E_8$, can give rise to a whole host of different subalgebras of $E_8$ depending on the value of $n_0$, according to \eqref{eq:massEntail}. This expectation is indeed borne out, in a sense we now explain.

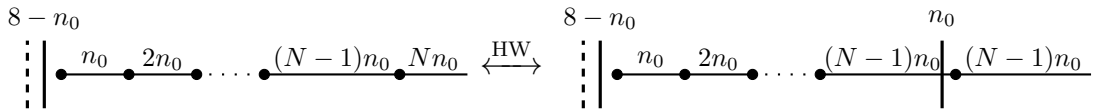
\begin{figure}[ht!]
\centering
\begin{tikzpicture}[scale=0.9,baseline]
\node at (0,0) {};
\draw[fill=black] (0.5,0) circle (0.075cm);
\draw[fill=black] (1.5,0) circle (0.075cm);
\draw[fill=black] (2.5,0) circle (0.075cm);
\draw[fill=black] (3.5,0) circle (0.075cm);
\draw[fill=black] (5.5,0) circle (0.075cm);


\draw[solid,black,thick] (0.5,0)--(1.5,0) node[black,midway,yshift=0.2cm] {\footnotesize $n_0$};
\draw[solid,black,thick] (1.5,0)--(2.5,0) node[black,midway,yshift=0.2cm] {\footnotesize $2n_0$};
\draw[loosely dotted,thick,black] (2.5,0)--(3.5,0) node[black,midway] {};
\draw[solid,black,thick] (3.5,0)--(5.5,0) node[black,midway,yshift=0.2cm] {\footnotesize $(N-1)n_0$};
\draw[solid,black,thick] (5.5,0)--(6.5,0) node[black,midway,xshift=0cm,yshift=0.2cm] {\footnotesize $Nn_0$};

\draw[dashed,black,very thick] (0,-.5)--(0,.5) node[black,midway, xshift =0cm, yshift=-1.5cm] {} node[black,midway, xshift =0cm, yshift=1.5cm] {};
\draw[solid,black,very thick] (0.25,-.5)--(0.25,.5) node[black,midway, xshift =0cm, yshift=+.75cm] {\footnotesize $8-n_0$};
\end{tikzpicture}
$\xleftrightarrow{\rm HW}$
\begin{tikzpicture}[scale=0.9,baseline]
\node at (0,0) {};
\draw[fill=black] (0.5,0) circle (0.075cm);
\draw[fill=black] (1.5,0) circle (0.075cm);
\draw[fill=black] (2.5,0) circle (0.075cm);
\draw[fill=black] (3.5,0) circle (0.075cm);
\draw[fill=black] (5.5,0) circle (0.075cm);


\draw[solid,black,thick] (0.5,0)--(1.5,0) node[black,midway,yshift=0.2cm] {\footnotesize $n_0$};
\draw[solid,black,thick] (1.5,0)--(2.5,0) node[black,midway,yshift=0.2cm] {\footnotesize $2n_0$};
\draw[loosely dotted,thick,black] (2.5,0)--(3.5,0) node[black,midway] {};
\draw[solid,black,thick] (3.5,0)--(5.5,0) node[black,midway,xshift=-2, yshift=0.2cm] {\footnotesize $(N-1)n_0$};
\draw[solid,black,thick] (5.5,0)--(7.5,0) node[black,midway,xshift=0cm,yshift=0.2cm] {\footnotesize $(N-1)n_0$};

\draw[dashed,black,very thick] (0,-.5)--(0,.5) node[black,midway, xshift =0cm, yshift=-1.5cm] {} node[black,midway, xshift =0cm, yshift=1.5cm] {};
\draw[solid,black,very thick] (0.25,-.5)--(0.25,.5) node[black,midway, xshift =0cm, yshift=+.75cm] {\footnotesize $8-n_0$};
\draw[solid,black,very thick] (5.5-0.2,-.5)--(5.5-0.2,.5) node[black,midway, xshift =0cm, yshift=+.75cm] {\footnotesize $n_0$};
\end{tikzpicture}

\caption{\textbf{Left:} massive Type IIA configuration dual to the F-theory electric quiver in \eqref{eq:massive}.  \textbf{Right:} massless Type IIA configuration (orbi-instanton) with $k-n_0$ semi-infinite D6's obtained from the massive setup at the top via $n_0$ Hanany--Witten (HW) moves.}
\label{fig:massive}
\end{figure}

Consider the massive quiver \eqref{eq:massive} for simplicity.  Its corresponding Type IIA setup is drawn in figure \ref{fig:massive}. 
The rightmost NS5 has $(N-1)n_0=k-n_0$ D6's ending on it from the left and $k=Nn_0$ from the right, due to D6 charge conservation in presence of a Romans mass (or equivalently gauge anomaly cancellation in the electric quiver). We can trade a semi-infinite D6 on the right for a D6 ending on a  D8. Using standard Hanany--Witten moves we can then bring the D8 across the rightmost NS5, removing the D6 that ends on it altogether. Repeating this operation $n_0$ times we end up with a massless brane system with $8-n_0+n_0=8$ D8's (crossing the last finite segment of D6's) but only $k-n_0$ semi-infinite D6's, rather than the original $k$. 

We have learned that a massive $E_{1+(8-n_0)}$-string theory (with empty or nonempty $-1$, generalizing in the obvious way the above argument) is equivalent via $n_0$ Hanany--Witten moves to an orbi-instanton with order of the orbifold equal to $k-n_0$ and length of the plateau equal to one.  That is, 
\begin{equation}\label{eq:HWmass}
\text{massive:}\ \cdots\ \overset{\su{k-n_0}}{2} \ [\SU(k)]\ \xrightarrow{\text{HW}} \ \text{orbi-instanton:}\ \cdots\ \overset{\su{k-n_0}}{\underset{[N_\text{f}=n_0]}{2}} \ [\SU(k-n_0)]\ ,
\end{equation}
to go from a massive theory to the corresponding orbi-instanton (with any electric quiver), or vice versa. It is then clear that the massive E-string theories are a special subclass of the orbi-instantons.  It remains to understand how to constrain the $E_8$ Kac labels in order to identify \emph{only} the massive E-strings among the labels for $k-n_0$ (in fact some labels will correspond to orbi-instantons that cannot be related via Hanany--Witten moves to a massive E-string at $n_0$).  Moreover we observe the above trick is enough to fully determine the magnetic quiver of the massive E-strings given what we said in section \ref{sub:magorbi}, a fact which will be useful later on to construct Higgs branch RG flows.

To constrain Kac labels, we can proceed in two independent ways yielding, of course, the same conclusion (summarized in table \ref{tab:constrE8}). 
\begin{table}[ht!]
\centering
\renewcommand{\arraystretch}{1.25}
\begin{tabular}{lccc} 
flavor & $n_0$ & orbifold & \text{constraints on $E_8$ Kac labels of $k-n_0$} \\
\hline
$E_8$ & $1$ & $k-1$ & none\\ 
$E_7$ & $2$ & $k-2$ & $n_1=0$\\
$E_6$ & $3$ & $k-3$ & $n_1=n_2=0$\\
$E_5=D_5$ & $4$ & $k-4$ & $n_1=n_2=n_3=0$\\
$E_4=A_4$ & $5$ & $k-5$ & $n_1=n_2=n_3=n_4=0$\\
$E_3=A_2\oplus A_1$ & $6$ & $k-6$ & $n_1=n_2=n_3=n_4=n_5=0$\\
$E_2=A_1\oplus \uu{1}$ & $7$ & $k-7$ & $n_1=n_2=n_3=n_4=n_5=n_6=0$\\
$E_1=A_1 $ (O8$^-$) & $8$ & $k-8$ & $n_1=n_2=n_3=n_4=n_5=n_6=0$  and \\ &&& $\begin{cases} n_{4'}=0\ \cap n_{2'}>\tfrac{1}{2}n_{3'} \ \text{or}\\ n_{3'}=0 \end{cases}$ \\
$\tilde{E}_1=\uu{1}$ (O8$^*$ $+$ D8)& $8$ & $k-8$ & $n_1=n_2=n_3=n_4=n_5=n_6=n_{4'}=0$ and \\ &&& $n_{2'}<\tfrac{1}{2}n_{3'}$ \\
$E_0=\emptyset$ (O8$^*$)& $9$ & $k-9$ & $n_1=n_2=n_3=n_4=n_5=n_6=n_{4'}=n_{2'}=0$
\end{tabular}
\caption{Constrained $E_8$ Kac labels of $k-n_0$ identifying massive $E_{1+(8-n_0)}$-string theories with $k$ semi-infinite D6's. In the $n_0=1$ case of the massive $E_8$-string even though the labels themselves are not constrained, $k$ and $N$ are, since $k=Nn_0=N$.  $\tilde{E}_1$ has an extra D8 on top of the O8$^*$; ${E}_0$ does not. Satisfactorily, the conditions for $\tilde{E}_1$ contain those for $E_0$ as a subcase, which is to be expected since the former theory in 5d flows to the latter via mass deformation \cite{Morrison:1996xf}.}
\label{tab:constrE8}
\end{table}

Looking at the electric quiver of the orbi-instanton with empty $-1$ and $k-n_0=(N-1)n_0$ (obtained via \eqref{eq:HWmass}),
\begin{equation}
\label{eq:orbi-less}
[E_{1+(8-n_0)}]\ {1}\ \overset{\mathfrak{su}(n_0)}{2}\ \overset{\mathfrak{su}(2n_0)}{2} \cdots \overset{\mathfrak{su}((N-1)n_0)}{\underset{[N_\text{f}=n_0]}{2}}\  [\SU((N-1)n_0)]\ ,
\end{equation}
we see that the ranks of the $\mathfrak{su}$ gauge algebras increase along the ramp by at least $n_0$.\footnote{Because of D6 charge conservation, the difference between the numbers of D6's ending from the left and right on an NS5 must be equal to the Romans mass ``felt'' by the NS5, which by convention is given by the total D8 charge to the immediate left. Clearly if in the finite interval immediately preceding the NS5 there is one or more extra D8's, this charge changes, and to compensate for it the rank of the $\mathfrak{su}$ algebra must jump by more than just $n_0$.} It is then enough to use the algorithm in \cite[Sec. 3.2]{Mekareeya:2017jgc} to determine the constrained $E_8$ labels associated to theories of the form \eqref{eq:orbi-less}. Using notation as in \cite[Eq. (3.3)]{Mekareeya:2017jgc} we define $a_s$ to be the number of times a difference equal to $s$ appears in the electric quiver:\footnote{In massive quivers, this is the same as the Romans mass in the interval between the $i$-th and $(i+1)$-th NS5; see e.g.  \cite[Eq. (2.11)]{Apruzzi:2017nck}.}
\begin{equation}\label{eq:as}
a_s = \# \{ i\ \text{s.t.}\ r_{i+1} - r_{i} = s \}\ .
\end{equation}
In \cite[Eq. (3.6)]{Mekareeya:2017jgc} it is also shown that $a_i=n_i$ for $i=1,\ldots,6$, with $n_i$ the multiplicities of the unprimed Coxeter labels as in \eqref{eq:kaccoord}. That is, if all differences are at least equal to $n_j$ for $j=1,\ldots,6$ we simply impose that all multiplicities $n_i$ with $i<j$ vanish in the Kac label associated with the electric quiver for the orbi-instanton at $k-n_0$, i.e. \eqref{eq:orbi-less}. Here we see the differences are at least $n_0$.  

One may wonder what happens if we start splitting the stack of $n_0$ D8's and move them around compatibly with D6 charge conservation, as this would generate different jumps.  However this is not always possible if we \emph{fix} the length $N$ of the massive quiver once and for all, as to satisfy charge conservation sometimes we would need to lengthen the quiver and allow for $\mathfrak{su}$ algebras with even higher ranks than $Nn_0$ (which is impossible if we fix $N$). 

All in all, for $n_0=1,\ldots,6$ the above procedure yields the first six rows of table \ref{tab:constrE8}. When the difference is equal to $7,8,9$ instead (the latter case being realized when the O8$^-$ is replaced by a lone O8$^*$, see figure \ref{fig:IIAE0}) we have to be more careful.  Once again, following the algorithm on \cite[p.9]{Mekareeya:2017jgc} we arrive at the last three rows of the table. 
 
Alternatively,  another way to produce the constraints of table \ref{tab:constrE8} is to realize that the $\SU(k-n_0)$ global symmetry of the orbi-instanton must enhance to $\SU(k)$ if the $E_8$ label identifies a massive $E_{1+(8-n_0)}$-string rather than an orbi-instanton. Then, we can impose the enhancement at the level of 3d magnetic quiver: it is enough to require that certain nodes in it be balanced (i.e. $2N_\text{c}=N_\text{f}$).  In fact, generically, the balanced part of the magnetic quiver gives the full flavor symmetry at the fixed point \cite{Gaiotto:2008ak}.\footnote{Exceptions to this rule are known. Sometimes the balanced part is only a subalgebra of the full (enhanced) flavor \cite{Mekareeya:2017sqh} (see also  \cite{Gledhill:2021cbe} for further examples).} Constructing the magnetic quiver of \eqref{eq:orbi-less},  we readily see that it has the expected structure \eqref{eq:genericmagquiv}, and that the $1- \cdots - (k-n_0-1)$ portion of the left tail is already balanced, providing the expected $\SU(k-n_0)$ symmetry.  To impose that the latter enhances to the full $\SU(k)$ we must require that the balanced tail is actually part of a bigger connected subquiver with $k$ nodes which are all balanced.  This in turn constrains the number $N$ and also the choice of labels that give rise to such a structure.

Either way, the constraints we obtain can be neatly summarized, and have been recorded in table \ref{tab:constrE8}.

\subsection{A holographic check}
\label{sub:holo}

To reinforce our claim that any massive theory at $n_0$ is equivalent to an orbi-instanton at $k-n_0$ and a length-one plateau, we will now compute their holographic duals \cite{Apruzzi:2013yva}. These are conveniently encoded in a single polynomial function of one variable, $\alpha(z)$ \cite{Cremonesi:2015bld,Apruzzi:2017nck}.\footnote{The coordinate $z$ parameterizes the ``near-horizon'' of the direction along which the NS5's are arranged, progressively far away from the O8$^-$.} Therefore it is enough to show that any of the physical observables extracted from $\alpha$ which can be matched with the leading term of the dual field theory observable (at large $N$) has the same value for both massive and orbi-instanton theory.  One such observable is the $a$ conformal anomaly.

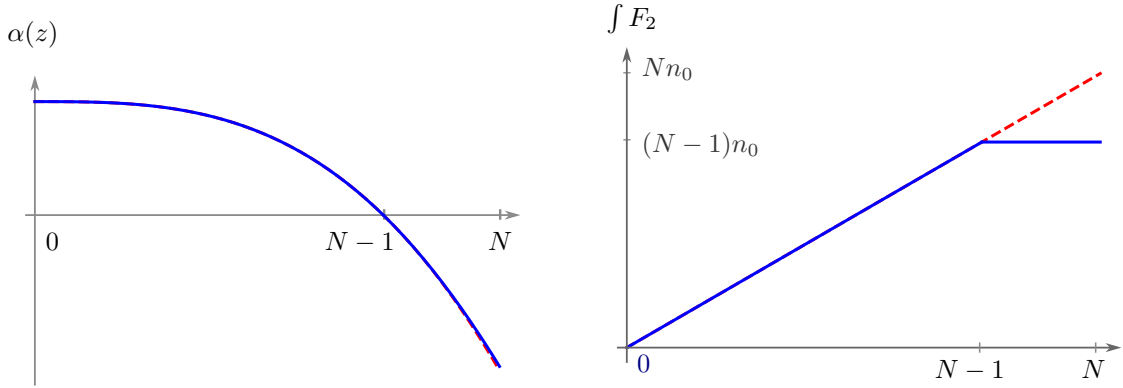
\begin{figure}
  \footnotesize{
  \centering
    \begin{subfigure}[b]{0.5\textwidth}
        \centering
        \def\svgwidth{\linewidth}
\begingroup%
  \makeatletter%
  \providecommand\color[2][]{%
    \errmessage{(Inkscape) Color is used for the text in Inkscape, but the package 'color.sty' is not loaded}%
    \renewcommand\color[2][]{}%
  }%
  \providecommand\transparent[1]{%
    \errmessage{(Inkscape) Transparency is used (non-zero) for the text in Inkscape, but the package 'transparent.sty' is not loaded}%
    \renewcommand\transparent[1]{}%
  }%
  \providecommand\rotatebox[2]{#2}%
  \newcommand*\fsize{\dimexpr\f@size pt\relax}%
  \newcommand*\lineheight[1]{\fontsize{\fsize}{#1\fsize}\selectfont}%
  \ifx\svgwidth\undefined%
    \setlength{\unitlength}{424.34335303bp}%
    \ifx\svgscale\undefined%
      \relax%
    \else%
      \setlength{\unitlength}{\unitlength * \real{\svgscale}}%
    \fi%
  \else%
    \setlength{\unitlength}{\svgwidth}%
  \fi%
  \global\let\svgwidth\undefined%
  \global\let\svgscale\undefined%
  \makeatother%
  \begin{picture}(1,0.62279524)%
    \lineheight{1}%
    \setlength\tabcolsep{0pt}%
    \put(0,0){\includegraphics[width=\unitlength,page=1]{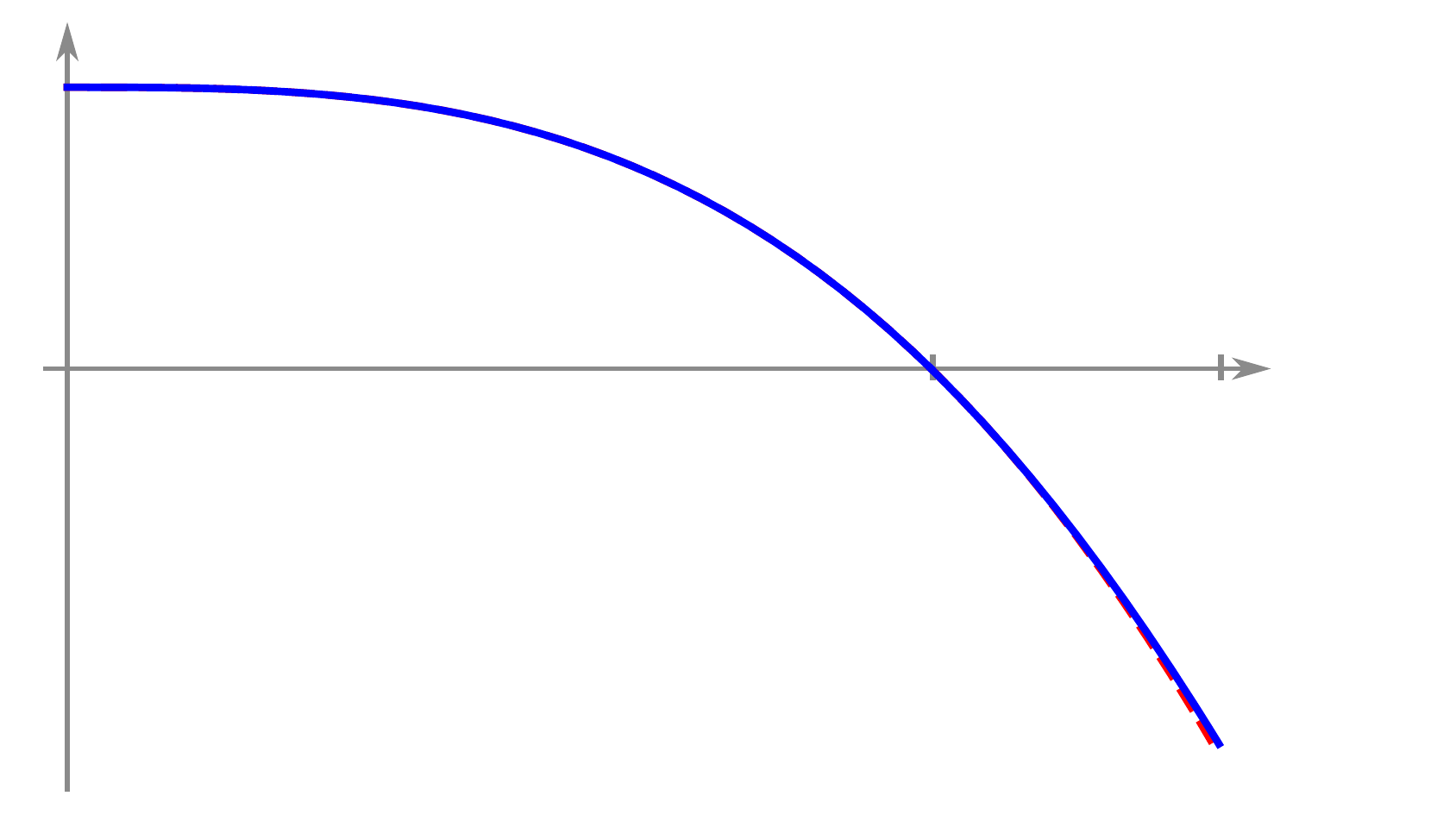}}%
    \put(-0.00022783,0.60709543){\color[rgb]{0,0,0}\makebox(0,0)[lt]{\lineheight{1.25}\smash{\begin{tabular}[t]{l}$\alpha(z)$\end{tabular}}}}%
    \put(0.06472403,0.2515095){\color[rgb]{0,0,0}\makebox(0,0)[lt]{\lineheight{1.25}\smash{\begin{tabular}[t]{l}$0$\end{tabular}}}}%
    \put(0.5401933,0.2515095){\color[rgb]{0,0,0}\makebox(0,0)[lt]{\lineheight{1.25}\smash{\begin{tabular}[t]{l}$N-1$\end{tabular}}}}%
    \put(0.81803195,0.2515095){\color[rgb]{0,0,0}\makebox(0,0)[lt]{\lineheight{1.25}\smash{\begin{tabular}[t]{l}$N$\end{tabular}}}}%
  \end{picture}%
\endgroup%

    \end{subfigure}%
    ~ 
    \begin{subfigure}[b]{0.5\textwidth}
        \centering
        \def\svgwidth{\linewidth}
\begingroup%
  \makeatletter%
  \providecommand\color[2][]{%
    \errmessage{(Inkscape) Color is used for the text in Inkscape, but the package 'color.sty' is not loaded}%
    \renewcommand\color[2][]{}%
  }%
  \providecommand\transparent[1]{%
    \errmessage{(Inkscape) Transparency is used (non-zero) for the text in Inkscape, but the package 'transparent.sty' is not loaded}%
    \renewcommand\transparent[1]{}%
  }%
  \providecommand\rotatebox[2]{#2}%
  \newcommand*\fsize{\dimexpr\f@size pt\relax}%
  \newcommand*\lineheight[1]{\fontsize{\fsize}{#1\fsize}\selectfont}%
  \ifx\svgwidth\undefined%
    \setlength{\unitlength}{425.90139323bp}%
    \ifx\svgscale\undefined%
      \relax%
    \else%
      \setlength{\unitlength}{\unitlength * \real{\svgscale}}%
    \fi%
  \else%
    \setlength{\unitlength}{\svgwidth}%
  \fi%
  \global\let\svgwidth\undefined%
  \global\let\svgscale\undefined%
  \makeatother%
  \begin{picture}(1,0.64916611)%
    \lineheight{1}%
    \setlength\tabcolsep{0pt}%
    \put(0,0){\includegraphics[width=\unitlength,page=1]{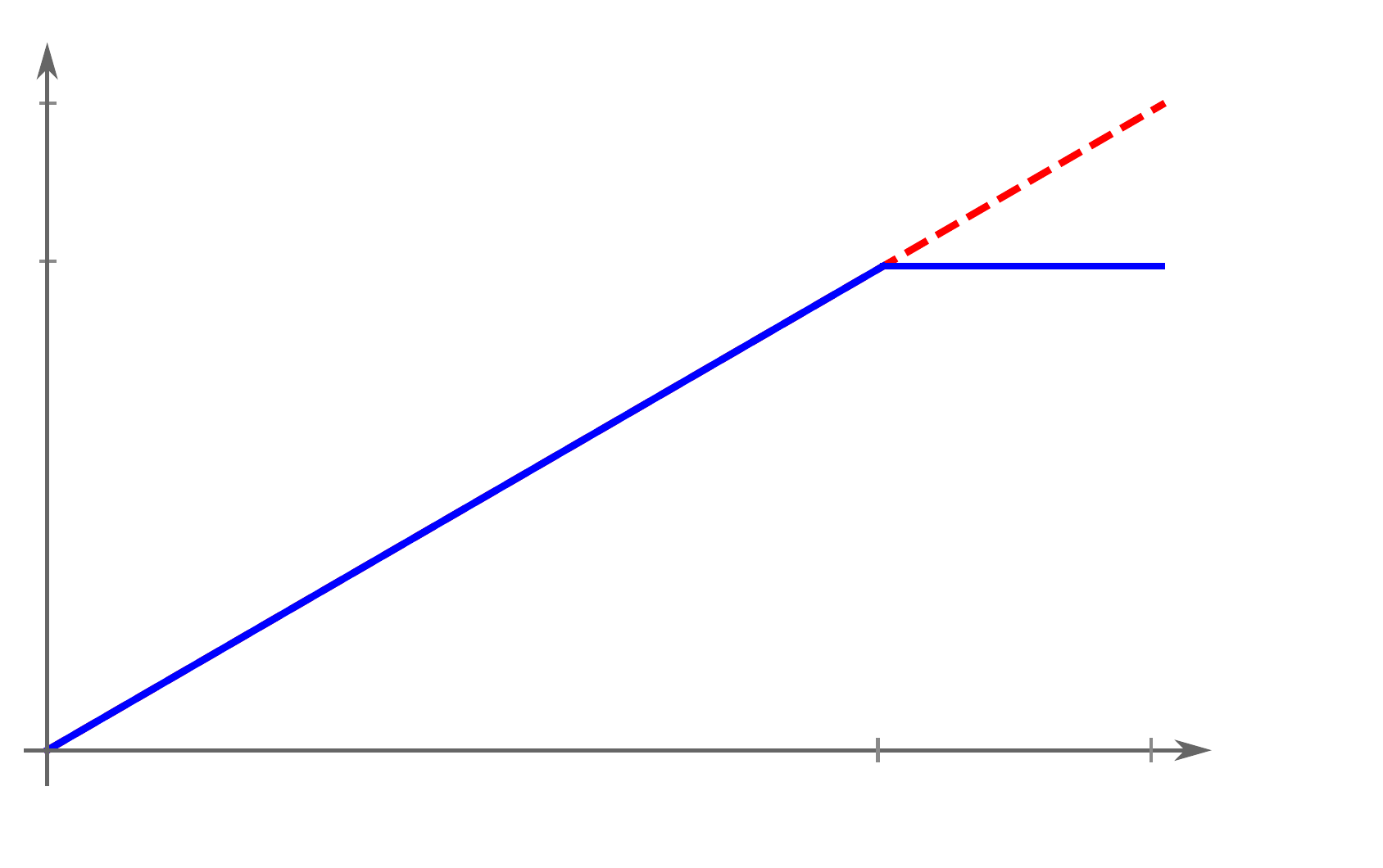}}%
    \put(-0.000227,0.63352373){\color[rgb]{0,0,0}\makebox(0,0)[lt]{\lineheight{1.25}\smash{\begin{tabular}[t]{l}$\int F_2$\end{tabular}}}}%
    \put(0.55884087,0.03273641){\color[rgb]{0,0,0}\makebox(0,0)[lt]{\lineheight{1.25}\smash{\begin{tabular}[t]{l}$N-1$\end{tabular}}}}%
    \put(0.8085508,0.03273641){\color[rgb]{0,0,0}\makebox(0,0)[lt]{\lineheight{1.25}\smash{\begin{tabular}[t]{l}$N$\end{tabular}}}}%
    \put(0.05169014,0.04330224){\color[rgb]{0,0,0.50196078}\makebox(0,0)[lt]{\lineheight{1.25}\smash{\begin{tabular}[t]{l}$0$\end{tabular}}}}%
    \put(0.05956005,0.4174097){\color[rgb]{0.2,0.2,0.2}\makebox(0,0)[lt]{\lineheight{1.25}\smash{\begin{tabular}[t]{l}$(N-1)n_0$\end{tabular}}}}%
    \put(0.05956005,0.54772155){\color[rgb]{0.2,0.2,0.2}\makebox(0,0)[lt]{\lineheight{1.25}\smash{\begin{tabular}[t]{l}$N n_0$\end{tabular}}}}%
  \end{picture}%
\endgroup%

    \end{subfigure}
  }
\caption{\textbf{Left:} $α(z)$ for the massive theory (red, dashed) and the corresponding orbi-instanton (blue, solid) obtained via Hanany--Witten moves. The two graphs overlap for $z\in [0,N]$ and diverge slightly in the interval $[N-1,N]$. \textbf{Right:} integral $\int F_2$ of the two-form RR flux  for the massive (red, dashed) and massless (blue, solid) theories.}
\label{fig:α-F2}
\end{figure}

Following the prescription in \cite{Apruzzi:2017nck}, it is easy to find that, at the two ends of the finite interval $I=[0,N]$ representing the base of the internal space $S^2 \hookrightarrow M_3 \to I$ of the dual AdS$_7$ vacua, we must enforce the following boundary conditions, representing O8-D8 and D6 sources:
\begin{align}
&\text{D8-O8 at $z=0$:}\quad  \dot{\alpha},\ddot{\alpha}\to 0\ , & &\text{D6 at $z = N$:}\quad  \alpha \to 0\ .
\end{align}
Enforcing these boundary conditions, the ``boundary data'' which are needed to construct the $\alpha$'s fall into the class of \cite[App. B.6]{Apruzzi:2017nck} (to which we refer the reader for more details on this construction) in both cases. With those, we can construct explicitly the $\alpha$'s. Focusing on the simplest case with empty $-1$ curve, we find:
\begin{equation}
\alpha(z)_\text{mass} = \frac{3^3 \pi^2}{2}  n_0 \left((N-1)^3-z^3\right)\ 
\end{equation}
for the massive $E_{1+(8-n_0)}$-string with $k=Nn_0$, and 
\begin{equation}
\alpha(z)_\text{orbi} = \begin{cases} \alpha(z)_\text{mass} & z\in [0,N-1] \\ \frac{3^4 \pi^2}{2} (N-1) n_0 z ((N-1)-z) & z\in [N-1,N] \end{cases}
\end{equation}
for the orbi-instanton at $k-n_0$ and with a plateau of length one.  We have plotted these functions in \fref{fig:α-F2}. Using $\alpha(z)$, we can compute for instance the RR flux $F_2$:
\begin{equation}
  \int_{S^2} F_2 = -\frac{1}{81π²} \ddot{α}\ .
\end{equation}
When integrated on the $S^2$ over the point $z=N$ this flux detects the number of D6 sources present at the right end of the supergravity dual. We can see from the right frame of figure \ref{fig:α-F2} that $F_2$ differs between massive and orbi-instanton string, as it should. Using $α(z)$ we can also compute the holographic $a$ anomaly \cite{Apruzzi:2015zna,Cremonesi:2015bld,Apruzzi:2017nck},
\begin{equation}
  a_\text{holo} = - \frac{128}{189 π²} \int_0^N \frac{α\ddot{α}}{2\left(9π\right)²} \, dz = \frac{48}{35}n_0^2 N^5 + \mathcal{O}\left(n_0² N^4\right)\ ,
\end{equation}
for both massive string and orbi-instanton, where the coefficients at order $N^4$ and lower are different, and are expected to be rendered equal only once one includes stringy and higher-derivative corrections to the supergravity solutions.  On the other hand, computing the exact $a$ anomaly for the massive string or orbi-instanton in figure \ref{fig:massive},\footnote{Since the two quiver gauge theories are connected by a Hanany--Witten move,  and are thus the same, we can use either to compute the $a$ exactly.} we obtain
\begin{equation}
a=\frac{16}{35} n_0^2 N\left(3 N^4-5 N^2+2\right)\ ,
\end{equation}
whose leading term at large $N$ matches the holographic $a$, as expected.

\section{\texorpdfstring{Higgs branch flows at fixed $n_0$}{Higgs branch flows at fixed n0}}
\label{sec:higgsflows}

\begin{figure}[ht!]
	\centering
\begin{tikzpicture}[scale=1,baseline]
	\node at (0,0) {};
	\draw[fill=black] (0.5,0) circle (0.075cm);
	\draw[fill=black] (1.5,0) circle (0.075cm);
	\draw[fill=black] (2.5,0) circle (0.075cm);
	\draw[fill=black] (3.5,0) circle (0.075cm);
	\draw[fill=black] (4.5,0) circle (0.075cm);
	
	
	\draw[solid,black,thick] (0.5,0)--(1.5,0) node[black,midway,yshift=0.2cm] {\footnotesize $2$};
	\draw[solid,black,thick] (1.5,0)--(2.5,0) node[black,midway,yshift=0.2cm] {\footnotesize $4$};
	\draw[solid,black,thick] (2.5,0)--(3.5,0) node[black,midway,yshift=0.2cm] {\footnotesize $6$};
	\draw[solid,black,thick] (3.5,0)--(4.5,0) node[black,midway,yshift=0.2cm] {\footnotesize $8$};
	\draw[solid,black,thick] (4.5,0)--(5.5,0) node[black,midway,yshift=0.2cm] {\footnotesize $10$};
	
	\draw[dashed,black,very thick] (0,-.5)--(0,.5) node[black,midway, xshift =0cm, yshift=-1.5cm] {} node[black,midway, xshift =0cm, yshift=1.5cm] {};
	\draw[solid,black,very thick] (0.25,-.5)--(0.25,.5) node[black,midway, xshift =0cm, yshift=+.75cm] {\footnotesize $6$};
\end{tikzpicture}
$\xleftrightarrow{\rm{HW}}$
\begin{tikzpicture}[scale=1,baseline]
	\node at (0,0) {};
	\draw[fill=black] (0.5,0) circle (0.075cm);
	\draw[fill=black] (1.5,0) circle (0.075cm);
	\draw[fill=black] (2.5,0) circle (0.075cm);
	\draw[fill=black] (3.5,0) circle (0.075cm);
	\draw[fill=black] (4.5,0) circle (0.075cm);
	
	
	\draw[solid,black,thick] (0.5,0)--(1.5,0) node[black,midway,yshift=0.2cm] {\footnotesize $2$};
	\draw[solid,black,thick] (1.5,0)--(2.5,0) node[black,midway,yshift=0.2cm] {\footnotesize $4$};
	\draw[solid,black,thick] (2.5,0)--(3.5,0) node[black,midway,yshift=0.2cm] {\footnotesize $6$};
	\draw[solid,black,thick] (3.5,0)--(4.5,0) node[black,midway,yshift=0.2cm] {\footnotesize $8$};
	\draw[solid,black,thick] (4.5,0)--(5.5,0) node[black,midway,yshift=0.2cm] {\footnotesize $8$};
	
	\draw[dashed,black,very thick] (0,-.5)--(0,.5) node[black,midway, xshift =0cm, yshift=-1.5cm] {} node[black,midway, xshift =0cm, yshift=1.5cm] {};
	\draw[solid,black,very thick] (0.25,-.5)--(0.25,.5) node[black,midway, xshift =0cm, yshift=+.75cm] {\footnotesize $6$};
	\draw[solid,black,very thick] (4.5-0.2,-.5)--(4.5-0.2,.5) node[black,midway, xshift =0cm, yshift=+.75cm] {\footnotesize $2$};
\end{tikzpicture}
$(n_0 = 2)$\\
\vspace*{-30pt}
$\Big\downarrow_{\rm Higgsing}$\\
\vspace*{-15pt}
\begin{tikzpicture}[scale=1,baseline]
	\node at (0,0) {};
	\draw[fill=black] (0.5,0) circle (0.075cm);
	\draw[fill=black] (1.5,0) circle (0.075cm);
	\draw[fill=black] (2.5,0) circle (0.075cm);
	\draw[fill=black] (3.5,0) circle (0.075cm);
	\draw[fill=black] (4.5,0) circle (0.075cm);
	
	
	\draw[solid,black,thick] (0.5,0)--(1.5,0) node[black,midway,yshift=0.2cm] {\footnotesize $2$};
	\draw[solid,black,thick] (1.5,0)--(2.5,0) node[black,midway,yshift=0.2cm] {\footnotesize $4$};
	\draw[solid,black,thick] (2.5,0)--(3.5,0) node[black,midway,yshift=0.2cm] {\footnotesize $6$};
	\draw[solid,black,thick] (3.5,0)--(4.5,0) node[black,midway,yshift=0.2cm] {\footnotesize $7$};
	\draw[solid,black,thick] (4.5,0)--(5.5,0) node[black,midway,yshift=0.2cm] {\footnotesize $7$};
	
	\draw[dashed,black,very thick] (0,-.5)--(0,.5) node[black,midway, xshift =0cm, yshift=-1.5cm] {} node[black,midway, xshift =0cm, yshift=1.5cm] {};
	\draw[solid,black,very thick] (0.25,-.5)--(0.25,.5) node[black,midway, xshift =0cm, yshift=+.75cm] {\footnotesize $6$};
	\draw[solid,black,very thick] (3.5-0.2,-.5)--(3.5-0.2,.5) node[black,midway, xshift =0cm, yshift=+.75cm] {\footnotesize $1$};
	\draw[solid,black,very thick] (4.5-0.2,-.5)--(4.5-0.2,.5) node[black,midway, xshift =0cm, yshift=+.75cm] {\footnotesize $1$};
\end{tikzpicture}
	$\xleftrightarrow{\rm{HW}}$
\begin{tikzpicture}[scale=1,baseline]
	\node at (0,0) {};
	\draw[fill=black] (0.5,0) circle (0.075cm);
	\draw[fill=black] (1.5,0) circle (0.075cm);
	\draw[fill=black] (2.5,0) circle (0.075cm);
	\draw[fill=black] (3.5,0) circle (0.075cm);
	\draw[fill=black] (4.5,0) circle (0.075cm);
	
	
	\draw[solid,black,thick] (0.5,0)--(1.5,0) node[black,midway,yshift=0.2cm] {\footnotesize $2$};
	\draw[solid,black,thick] (1.5,0)--(2.5,0) node[black,midway,yshift=0.2cm] {\footnotesize $4$};
	\draw[solid,black,thick] (2.5,0)--(3.5,0) node[black,midway,yshift=0.2cm] {\footnotesize $6$};
	\draw[solid,black,thick] (3.5,0)--(4.5,0) node[black,midway,yshift=0.2cm] {\footnotesize $7$};
	\draw[solid,black,thick] (4.5,0)--(5.5,0) node[black,midway,yshift=0.2cm] {\footnotesize $8$};
	
	\draw[dashed,black,very thick] (0,-.5)--(0,.5) node[black,midway, xshift =0cm, yshift=-1.5cm] {} node[black,midway, xshift =0cm, yshift=1.5cm] {};
	\draw[solid,black,very thick] (0.25,-.5)--(0.25,.5) node[black,midway, xshift =0cm, yshift=+.75cm] {\footnotesize $6$};
	\draw[solid,black,very thick] (3.5-0.2,-.5)--(3.5-0.2,.5) node[black,midway, xshift =0cm, yshift=+.75cm] {\footnotesize $1$};
\end{tikzpicture}
$(n_0 = 1)$
	\caption{The top row represents a massive E-string theory at $k=10, n_0 = 2$ which, via Hanany--Witten (HW) moves, is equivalent to an orbi-instanton at $k=8$. Higgsing the flavor group on the stack of 2 D8's results in the bottom row. Via HW moves, this is now equivalent to a massive E-string theory at $k=8, n_0=1$.}
	\label{fig:n0stack}
\end{figure}
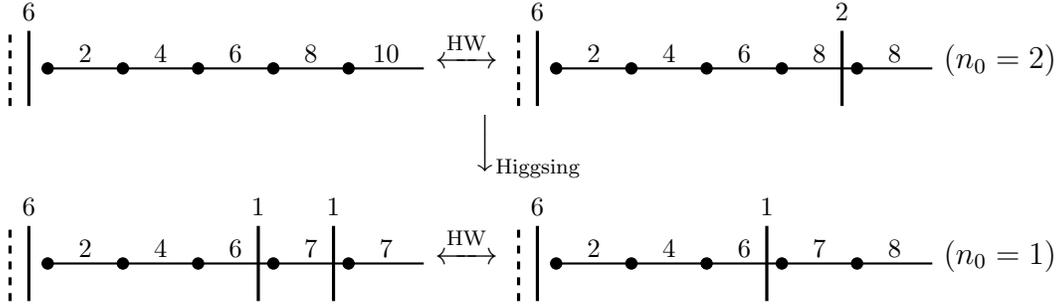

In this section we construct RG flows between massive E-strings at \emph{fixed} $n_0$.  This is the case that presents the least novelty, since it is a close copy of what happens for orbi-instantons \cite{Giacomelli:2022drw,Fazzi:2022hal}.  Once we fix the class of massive strings between which we wish to construct flows,  we give vev's to the moment maps of $E_{1+(8-n_0)}$ and flow between constrained $E_8$ Kac labels at fixed $n_0$.  That is, the flows between massive theories at $n_0$ are a subset of the flows between orbi-instantons at $k-n_0$.  For this reason, we can use the technology set up in \cite{Giacomelli:2022drw,Fazzi:2022hal} to construct the flows, i.e. 3d magnetic quivers and quiver subtraction between these.  As for ``unconstrained'' orbi-instantons,  all flows can be understood in the massive IIA brane setup as peeling off D8-branes from the stack on the O8.  We exclude flows corresponding to subdivisions of the stack of $n_0$ D8's as this would constitute a flow that changes $n_0$ (as shown in the example of \fref{fig:n0stack}).  We will devote section \ref{sec:Tflows} to this latter type of flows.

One remark is in order. In \cite{Fazzi:2022hal} we decided to fix $N$ to construct flows between orbi-instantons for clarity of exposition.  Doing so produces flows where the Kac label $[1^k]$, which preserves the full $E_8$ for any $k$, is always at the most \emph{bottom} position in the hierarchy. However one may naively expect it to be at the \emph{top} instead, since in a sense it is the largest flavor symmetry $\mathfrak{f}$ preserved by any label. This expectation is indeed borne out if one allows $N$ to change in the flows via small instanton transition (corresponding to the addition of a full $E_8$ Dynkin diagram to the right tail of the magnetic quiver, see \eqref{eq:genericmagquiv}). A clear example of this can be seen comparing \cite[Fig. 11(c)]{Fazzi:2022hal} and \cite[Fig. 12]{Fazzi:2022hal}, which give the hierarchy of RG flows for the $k=4$ A-type orbi-instanton at fixed $N$ and changing $N$, respectively. In \cite{fazzi-giri-levy} it has been conjectured that the two graphs correspond to two possible ``slicings'' of the so-called double affine Grassmannian of $E_8$, whose Hasse diagram of $E_8$ orbits is parameterized both by the labels and by $N$.

For massive theories, seen as orbi-instantons at $k-n_0$ with constrained Kac labels, we are \emph{forced} to allow $N$ to change to construct the RG flows.  An easy way to see this is as follows. If we fix $n_0$, we are also fixing the highest rank in the length-one plateau of the corresponding orbi-instanton, $k-n_0$. This also means the number of semi-infinite D6's to the right is fixed.  When we flow between constrained Kac labels, the ranks of the electric quivers associated to each of them jump by at least $n_0$. As a consequence, $N$ (i.e. the total number of compact curves, or gauge algebras, possibly including a trivial one with zero rank) is forced to get reduced along the flow so that the rank of the last gauge algebra does not ``overshoot'' $k-n_0$, which was fixed by hypothesis.  We will indicate the value of $N$ for a massive theory by a subscript on the constrained $E_8$ Kac labels for $k-n_0$.

\subsection{\texorpdfstring{Some examples for $n_0=1,\ldots,9$}{Some examples for n0=1,...,9}}
Here below we present some examples of flows for all possible values of $n_0$.  There are a few observations worth making.
\begin{itemize}
\item All edges in these hierarchies are labeled by the corresponding quiver subtraction, i.e. a Kraft--Procesi transition of type $A, \mathfrak{a},\mathfrak{d},\mathfrak{e}$. This is because the singularities of the $E_8$-orbits of the (double) affine Grassmannian of $E_8$ (which conjecturally gives the Higgs branch of the orbi-instanton \cite{Fazzi:2022hal}) can only be of those types \cite[Thm. A]{malin-ostrik-vybornov}. 
\item In some of the examples (see figures \ref{fig:k12n78} and \ref{fig:k30n89}) we have encountered isolated theories, i.e. IR SCFTs defined by allowed constrained Kac labels which are not reached by (or are not starting points of) any flows.  
\item Some of the allowed constrained labels in the hierarchies are related by a 6d $\theta$ angle (e.g.  $[2'^2,3'^2]_{\theta=\pi}$ and $[4'^2,2']_{\theta=0}$ for $k-n_0=10$ in \fref{fig:n23}).  We have colored in red the occurrences of SCFTs belonging to a $\theta$ angle ``pair''.  This is likely to occur in other examples (i.e. other values of $n_0$) for sufficiently high $k$, given the genericity of these so-called ``parallel flows'' \cite[Sec. 4.3]{Fazzi:2022hal}. 
\item It is quite satisfactory to note that the $k=5, n_0=1$ case (i.e. the massive $E_8$-string with $k=5$, equivalent to the orbi-instanton at $k'=5-1=4$) matches what originally found (via other techniques) in \cite[Fig. 1]{Frey:2018vpw}, which is reproduced in \cite[Fig. 12]{Fazzi:2022hal} via magnetic quiver subtraction. 
\end{itemize}

\begin{figure}[ht!]
\centering
\begin{subfigure}{0.45\textwidth}
\centering
\includegraphics[height=0.9\textheight]{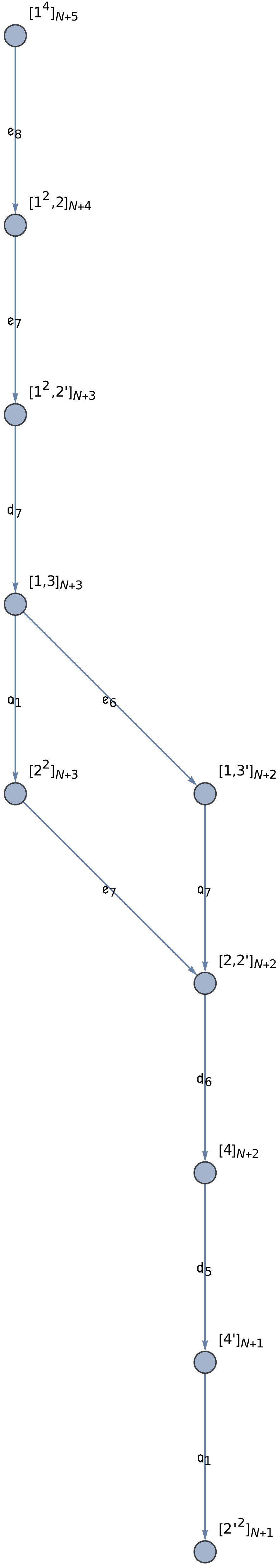}
\caption{}
\end{subfigure}
\begin{subfigure}{0.5\textwidth}
	\centering
  \includegraphics[height=0.9\textheight]{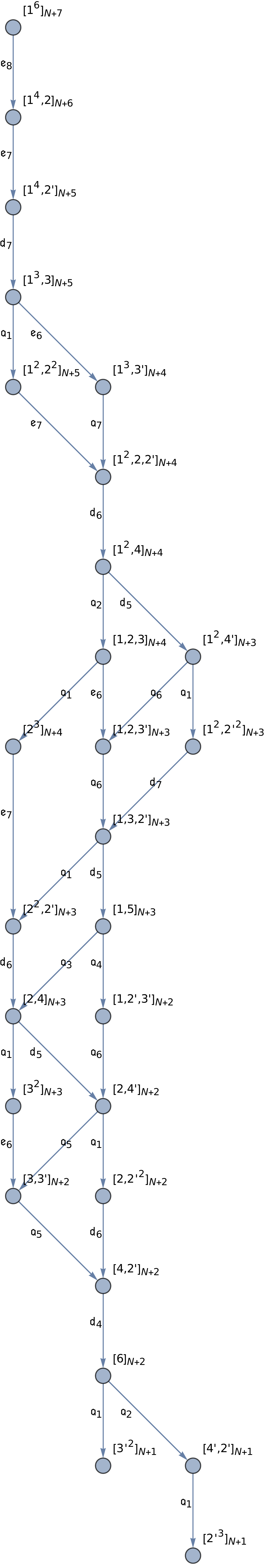}
  \caption{}
\end{subfigure}
\caption{Hierarchy of RG flows for \textbf{\emph{(a)}} massive $E₈$-theories with $n_0=1$ and $k=5$ (i.e. orbi-instanton at $k'=k-n_0=5-1=4$); \textbf{\emph{(b)}}  massive $E₈$-theories with $n_0=1$ and $k=7$ (i.e. orbi-instanton at $k=7-1=6$).}
\label{fig:n1}
\end{figure}
\begin{figure}[ht!]
	\centering
	\begin{subfigure}{0.45\textwidth}
		\centering
		\includegraphics[height=0.9\textheight]{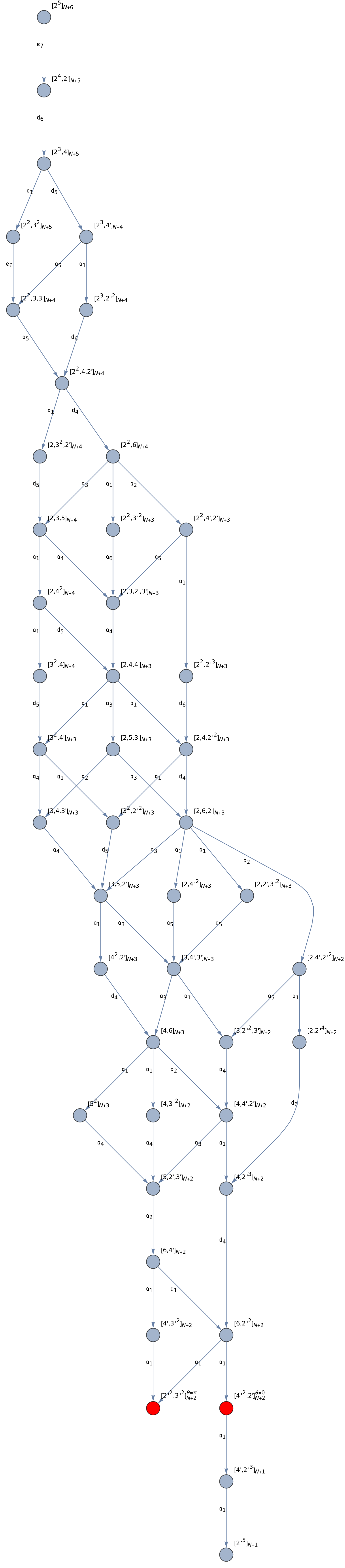}
		\caption{}
	\end{subfigure}
	\begin{subfigure}{0.5\textwidth}
		\centering
		\includegraphics[height=0.9\textheight]{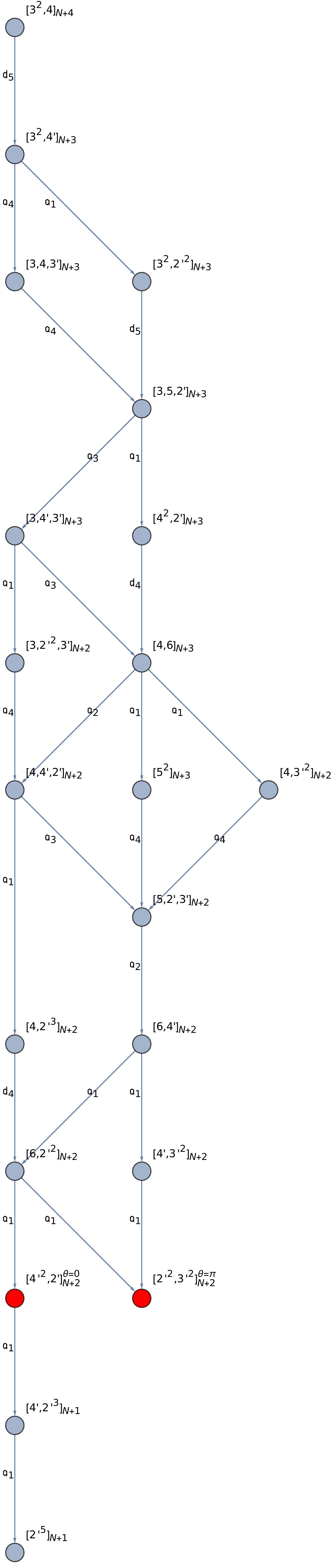}
		\caption{}
	\end{subfigure}
	\caption{Hierarchy of RG flows for \textbf{\emph{(a)}} massive $E₇$-theories with $n_0=2$ and $k=12$, and \textbf{\emph{(b)}} massive $E_6$-theories with $n_0=3$ and $k=13$. The labels colored in red differ by a 6d $\theta$ angle along the tensor branch.}
	\label{fig:n23}
\end{figure}
%
%

\begin{figure}[ht!]
	\centering
	\includegraphics[height=0.9\textheight]{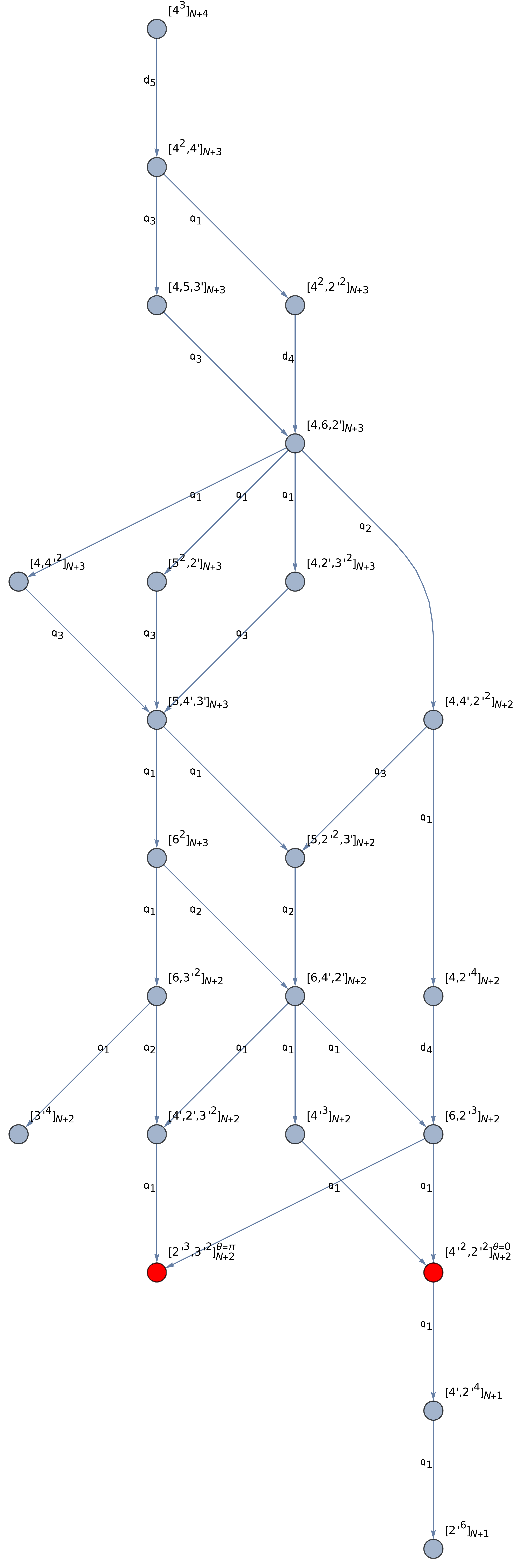}
	\caption{Hierarchy for massive $E_5$-theories with $n_0=4$ and $k=16$. }
	\label{fig:n4}
\end{figure}

\begin{figure}[ht!]
	\centering
	\includegraphics[height=0.9\textheight]{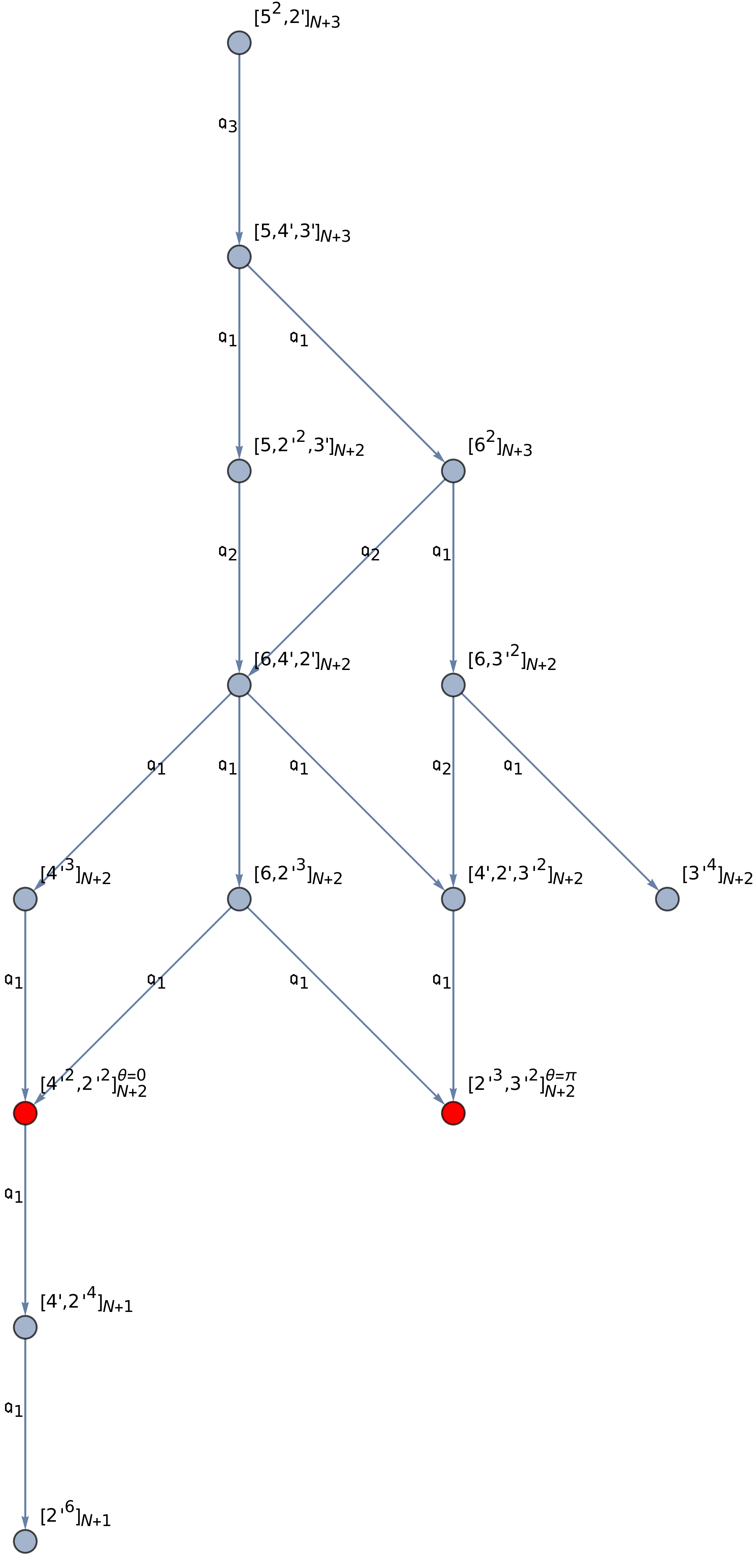}
	\caption{Hierarchy for massive $E_4$-theories with $n_0=5$ and $k=17$. }
	\label{fig:n5}
\end{figure}

\begin{figure}[ht!]
	\centering
	\includegraphics[height=0.9\textheight]{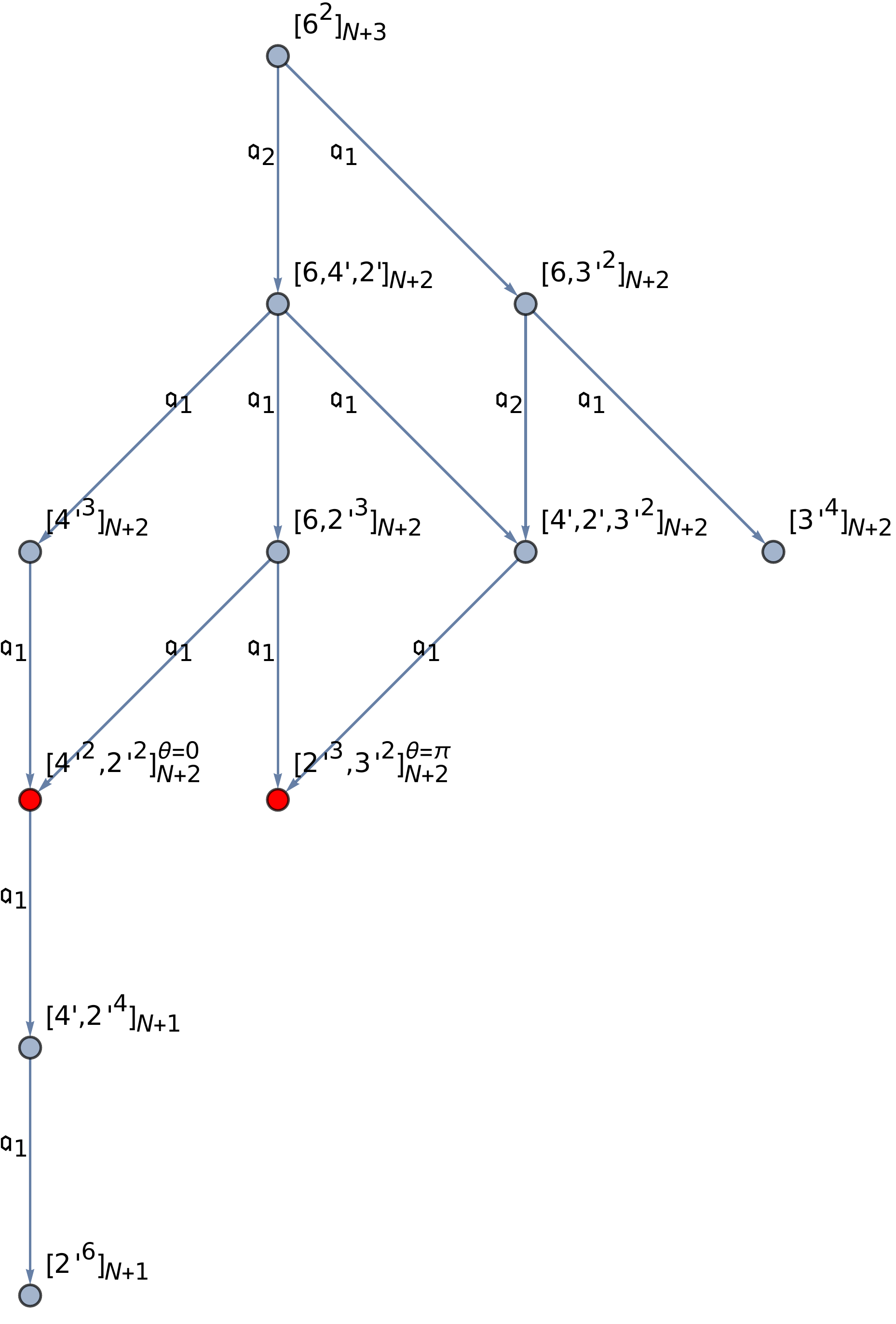}
	\caption{Hierarchy for massive $E_3$-theories with $n_0=6$ and $k=18$. }
	\label{fig:n6}
\end{figure}

\begin{figure}[ht!]
	\begin{subfigure}{0.5\textwidth}
		\centering
		\includegraphics[height=0.5\textheight]{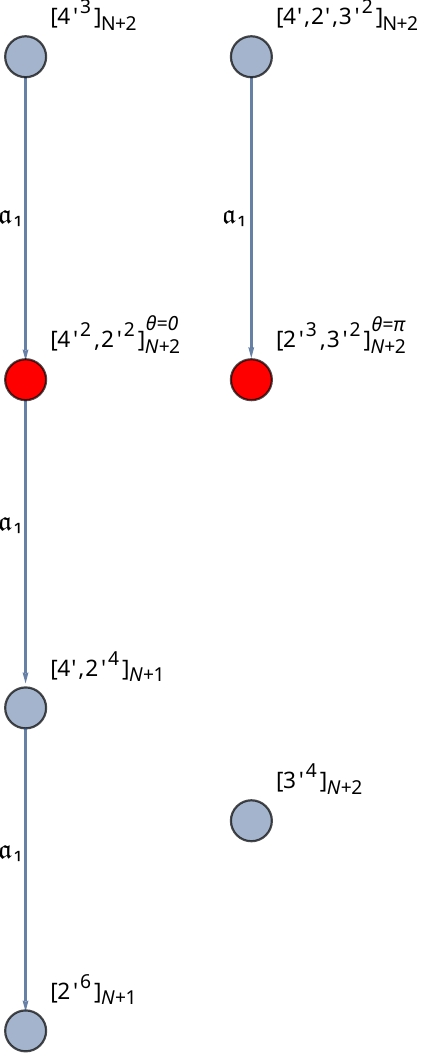}
		\caption{}
	\end{subfigure}
	\hfill
	\begin{subfigure}{0.5\textwidth}
		\centering
		\includegraphics[height=0.5\textheight]{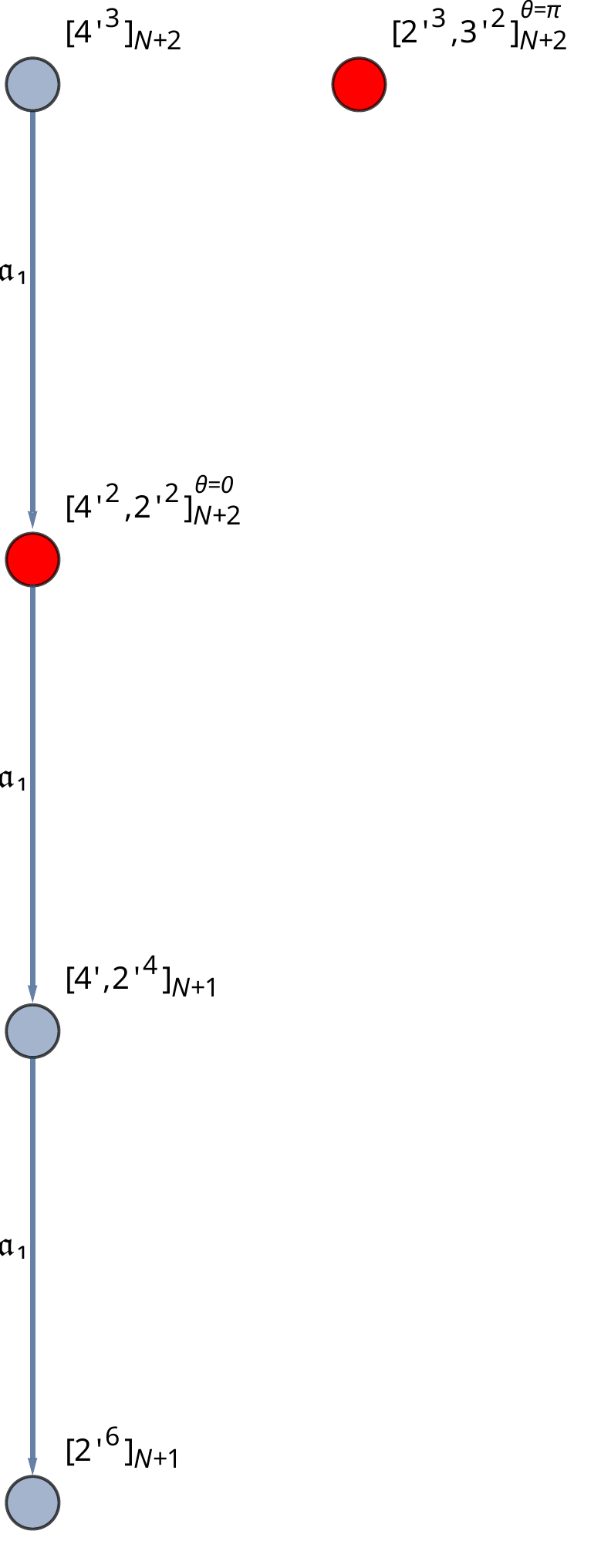}
		\caption{}
	\end{subfigure}

	\caption{\textbf{\emph{(a)}} Hierarchy for massive $E_2$-theories with $n_0=7$ and $k=19$; \textbf{\emph{(b)}} hierarchy for massive $E_1$-theories with $n_0=8$ and $k=20$.  Notice the presence of an isolated theory in both cases.}
	\label{fig:k12n78}
\end{figure}

\begin{figure}[ht!]
	\begin{subfigure}{0.5\textwidth}
		\centering
		\includegraphics[width=0.2\linewidth]{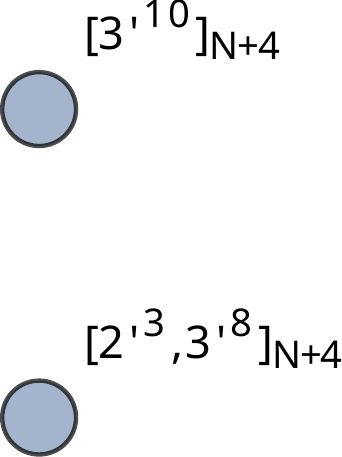}
		\caption{}
	\end{subfigure}
	\hfill
	\begin{subfigure}{0.5\textwidth}
		\centering
		\includegraphics[width=0.2\linewidth]{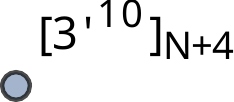}
		\caption{}
	\end{subfigure}
	
	\caption{\textbf{\emph{(a)}} Hierarchy for massive $\tilde{E}_1$-theories with $n_0=8$ and $k=38$; there are two isolated theories. \textbf{\emph{(b)}} Hierarchy for massive $E_0$-theories with $n_0=9$ and $k=39$; there is a single isolated theory.}
	\label{fig:k30n89}
\end{figure}

\clearpage

\section{\texorpdfstring{Higgs branch T-brane flows to different $n_0$}{Higgs branch T-brane flows to to different n0}}
\label{sec:Tflows}

In this section we will be less systematic, and explore some examples of a new type of RG flow that can be triggered in massive theories.

As we have seen all massive theories associated with a Type IIA brane system with $k$ semi-infinite D6-branes are equivalent to a massless theory with orbifold order $k-n_0$ and a length-one plateau. However, contrary to \emph{generic} $k-n_0$ orbi-instantons, these must have an enhanced $\SU(k)$ symmetry and we will now study the effect of a nilpotent vev for the $\SU(k)$ moment map. Nilpotent vev's for the $\SU(k-n_0)$ global symmetry (which exists for all orbi-instantons) have already been considered in the literature,  and have been identified with T-brane configurations in F-theory for the non\hyp{}perturbative seven-branes wrapping the noncompact $-2$ curve hosting this flavor algebra \cite{Gaiotto:2014lca,DelZotto:2014hpa,Heckman:2016ssk}.  Here we would like to point out that the enhanced global symmetry of the massive theories leads to new effects and the purpose of this section is to discuss in detail this aspect.  Such vev's have also been considered in \cite[Sec. 5.6]{Mekareeya:2017jgc} but only for generic orbi-instantons with sufficiently long plateau, so that their effects do not propagate to the far left. Here we want to achieve the opposite.

We can approach this problem in two closely related ways: either using the brane setup, by letting multiple semi-infinite D6's  end on a \emph{single} D8, or by modifying the magnetic quiver in suitable ways.  As we will see, these flows do ``mix'' left $E_{1+(8-n_0)}$ and right $\SU(k)$ flavor factors, a situation which has been vastly avoided in the literature because of its complication (with respect to vev's for decoupled factors). On the other hand, employing the 3d magnetic quiver technology allows us to deal with this mixing quite easily: the vev for the $\SU(k)$ factor propagates to the rest of the magnetic quiver, unbalancing some of the nodes, and thus modifying the preserved flavor algebra to a common subalgebra of $E_{1+(8-n_0)} \oplus \su{k}$. Rebalancing appropriately we land on the magnetic quiver for a new massive string at a \emph{different} value of $n_0$.

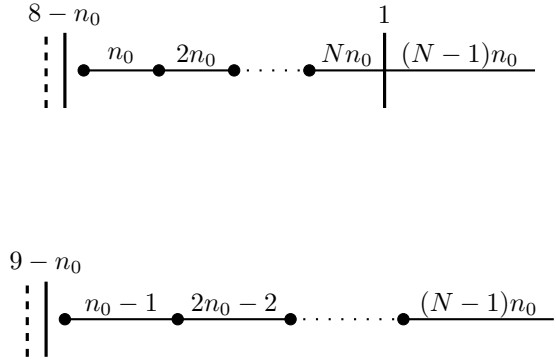
\begin{figure}[ht!]
\centering

\begin{tikzpicture}[scale=1,baseline]
\node at (0,0) {};
\draw[fill=black] (0.5,0) circle (0.075cm);
\draw[fill=black] (1.5,0) circle (0.075cm);
\draw[fill=black] (2.5,0) circle (0.075cm);
\draw[fill=black] (3.5,0) circle (0.075cm);


\draw[solid,black,thick] (0.5,0)--(1.5,0) node[black,midway,yshift=0.2cm] {\footnotesize $n_0$};
\draw[solid,black,thick] (1.5,0)--(2.5,0) node[black,midway,yshift=0.2cm] {\footnotesize $2n_0$};
\draw[loosely dotted,thick,black] (2.5,0)--(3.5,0) node[black,midway] {};
\draw[solid,black,thick] (3.5,0)--(4.5,0) node[black,midway,xshift=0cm,yshift=0.2cm] {\footnotesize $Nn_0$};
\draw[solid,black,thick] (4.5,0)--(6.5,0) node[black,midway,xshift=-0cm,yshift=0.2cm] {\footnotesize $(N-1)n_0$};

\draw[dashed,black,very thick] (0,-.5)--(0,.5) node[black,midway, xshift =0cm, yshift=-1.5cm] {} node[black,midway, xshift =0cm, yshift=1.5cm] {};
\draw[solid,black,very thick] (0.25,-.5)--(0.25,.5) node[black,midway, xshift =0cm, yshift=+.75cm] {\footnotesize $8-n_0$};
\draw[solid,black,very thick] (4.5,-.5)--(4.5,.5) node[black,midway, xshift =0cm, yshift=+.75cm] {\footnotesize $1$};
\end{tikzpicture}

\begin{tikzpicture}[scale=1,baseline]
\node at (0,0) {};
\draw[fill=black] (0.5,0) circle (0.075cm);
\draw[fill=black] (2,0) circle (0.075cm);
\draw[fill=black] (3.5,0) circle (0.075cm);
\draw[fill=black] (5,0) circle (0.075cm);


\draw[solid,black,thick] (0.5,0)--(2,0) node[black,midway,yshift=0.2cm] {\footnotesize $n_0-1$};
\draw[solid,black,thick] (2,0)--(3.5,0) node[black,midway,yshift=0.2cm] {\footnotesize $2n_0-2$};
\draw[loosely dotted,thick,black] (3.5,0)--(5,0) node[black,midway] {};
\draw[solid,black,thick] (5,0)--(7,0) node[black,midway,xshift=0cm,yshift=0.2cm] {\footnotesize $(N-1)n_0$};

\draw[dashed,black,very thick] (0,-.5)--(0,.5) node[black,midway, xshift =0cm, yshift=-1.5cm] {} node[black,midway, xshift =0cm, yshift=1.5cm] {};
\draw[solid,black,very thick] (0.25,-.5)--(0.25,.5) node[black,midway, xshift =0cm, yshift=+.75cm] {\footnotesize $9-n_0$};
\end{tikzpicture}

\caption{\textbf{Top:} activating one Jordan block of size $N$ means letting $N$ (out of $Nn_0$) D6's end on a single D8.  \textbf{Bottom:} we can then rearrange the brane system via simple Hanany--Witten moves.}
\label{fig:jordan1}
\end{figure}

First, consider for simplicity the case with empty $-1$ curve.  As done in figure \ref{fig:massive}, we can replace the $k=Nn_0$ semi-infinite D6-branes with $k$ D6's each ending on a different D8.  To activate a nilpotent vev for the $\SU(k)$ symmetry rotating them simply means letting multiple D6's end on the \emph{same} D8 \cite{Gaiotto:2008ak}, where the number of D6's ending on each D8 encodes the size of the Jordan blocks of the nilpotent vev matrix. We should then take into account the fact that the S-rule forces us to suspend at most one D6 between an NS5 and a D8. As a result, if we activate a vev with a single Jordan block of size $N$ in the brane system of figure \ref{fig:massive}, we find figure \ref{fig:jordan1}.
It is a simple exercise to bring in more D8's with further Hanany--Witten moves and their final location (assuming there are no D6's ending on them) depends on the size of the corresponding Jordan blocks. In particular, we end up with a massless configuration if we bring in exactly $n_0$ D8-branes (which is always possible since $k=Nn_0>n_0$ by construction). 

The corresponding orbi-instanton has orbifold order given by
\begin{equation}
k’=k-\sum_{i=1}^{n_0}N_i\ ,
\end{equation}
where $N_i$ denotes the size of the $i$-th Jordan block (i.e. the $i$-th D8 has $N_i$ D6's ending on it). When $N_i=1$ for all $i$ there is no nilpotent vev, the RG flow is trivial and we recover the correspondence between massive and massless theories we have discussed in section \ref{sub:constrainedkac}.  So in a sense massive strings at $n_0$ and orbi-instantons at $k-n_0$ are connected by a trivial flow, which we can generalize by activating nontrivial T-brane vev's. In fact by activating a nilpotent Higgsing for the $\SU(k)$ symmetry we can obtain for every massive theory a collection of several massless models, labeled by different orbifold orders. This feature is \emph{not} present for generic orbi-instantons, where typical Higgs branch RG flows do not change the orbifold order (e.g. this is held fixed in references \cite{Giacomelli:2022drw, Fazzi:2022hal}).

We will analyze in turn the three different classes of electric quivers mentioned at the end of section \ref{sub:massive} (i.e.  $1$,  $\overset{\mathfrak{usp}}{1}$ or $\overset{\mathfrak{su}}{1}$). Let us state our general result here and give an explanation by means of examples in the three following subsections (one per class).  As long as the size of all Jordan blocks does not exceed the number $N$ of NS5-branes, the gauge algebra supported on the $-1$ curve  is not Higgsed and the effect of the vev can be written in terms of the jumps in ranks $r_i$ of the algebras supported on consecutive $-2$ curves. The result is that for a Jordan block of size $N_i$ the last $N_i-1$ gauge algebras are Higgsed nontrivially and their rank differences $r_{i+1}-r_i$ decrease by one unit.  If instead  the size of one Jordan block is equal to the number of NS5's plus one, the rank of the gauge algebra supported on the $-1$ curve decreases by one unit. If that algebra is $\mathfrak{su}$ it also acquires one fundamental hypermultiplet in the process, whereas if it is $\mathfrak{usp}$ the number of fundamentals stays the same, but the number of fundamentals for the second gauge algebra increases by one unit. 

Let us analyze nilpotent vev's for the theory we will encounter in \eqref{brane5} to illustrate how our general result works in practice. The gauge algebras are $\su{4}$, supported on the $-1$ curve, $\su{7}$ and $\su{10}$ and therefore the ranks jump by three each time. In terms of the $a_s$ parameters introduced in \eqref{eq:as} we therefore have $a_3=2$ and all other $a_s$'s vanish. 
The corresponding constrained Kac label is $[3^2,4’]$ and when we turn on a Jordan block of size two $\su{4}$ and $\su{7}$ survive whereas $\su{10}$ is Higgsed to $\su{9}$. As a result we now have $a_2=a_3=1$ and the Kac label becomes $[2,3,4’]$. With a block of size three $\su{10}$ is higgsed to $\su{8}$ and $\su{7}$ to $\su{4}$, hence $a_3$ becomes zero and $a_2=2$. In terms of Kac labels we find $[2^2,4’]$, while we get $[2^2,3’]$ in the case of a block of size four, when the gauge group reduces to $\su{3}\oplus \su{5}\oplus \su{7}$. In the last step we see the Higgsing of the gauge algebra supported on the $-1$ curve. If in the latter case we then turn on a second Jordan block of e.g. size two we get $[1,2,3’]$ since the $\su{7}$ algebra is further Higgsed to $\su{6}$, whereas a second block of size three leads to $[1^2,3’]$, i.e. the gauge group becomes $\su{3}\oplus \su{4}\oplus\su{5}$. One can easily implement this process in the general case in a straightforward manner.

\subsection{\texorpdfstring{Electric quivers with empty $-1$ curve}{Electric quivers with empty -1 curve}}

We have just seen how to analyze the effect of a nilpotent vev for theories with an empty $-1$ curve  at the level of branes in figure \ref{fig:jordan1}. The same result can be obtained by implementing the Higgsing at the level of the corresponding magnetic quiver. In order to explain how this works, we have to briefly recall the properties of $T(\SU(N))$ \cite{Gaiotto:2008ak}. This is a 3d $\Nfour$ quiver gauge theory of the form 
\begin{equation}
\label{eq:tsu} 
1-2-\cdots-(N-2)-(N-1)-\boxed{N}
\end{equation} 
adopting the same notation as in \eqref{eq:genericmagquiv}, where moreover $\boxed{N}$ denotes a flavor node.  The theory has a $\U(1)^{N-1}$ topological symmetry which enhances to $\SU(N)$ due to the fact that all the gauge nodes are balanced. If we now turn on a nilpotent vev for the $\SU(N)$ moment map with Jordan block of size $n_1\geq n_2\geq \dots \geq n_{\ell}$, the $T(\SU(N))$ theory flows in the IR to 
\begin{equation}
\label{eq:niltsu}
\cdots - (N-n_1-n_2) - (N-n_1) -  \boxed{N}\ .
\end{equation} 
This is relevant for us because the magnetic quiver associated with our massive theories is given by a $T(\SU(k))$ theory coupled to a quiver with the shape of th $E_{1+(8-n_0)}$ Dynkin diagram and we can therefore analyze the effect of the nilpotent vev for $\SU(k)$ by exploiting the result we just recalled in \eqref{eq:niltsu}.  

Consider for instance the massive theory at $n_0=2$ with the following electric quiver, characterized by an empty $-1$ curve (i.e. no D6's cross the O8$^-$):
\begin{equation}
\label{eq:ex1}
k-2=2N-2=[2^{N-1}]:\ [E_{7}]\ 1\ \overset{\mathfrak{su}(2)}{2}\ \overset{\mathfrak{su}(4)}{2}\ \overset{\mathfrak{su}(6)}{2} \cdots \ \overset{\mathfrak{su}(2N-2)}{2} \ [\SU(2N)]\ ,
\end{equation}
the corresponding magnetic quiver is given by 
\begin{equation}\label{eq:ex1mag}
 {
\underbrace{ 1 - 2 - \cdots - (2N-1)}_{T(\SU(2N))} - 2N-\underbrace{ 3N -4N -5N-\overset{\overset{\displaystyle 3N}{\vert}}{6N}-4N-2N}_{\text{$E_7$ Dynkin}}}
\end{equation}
If we turn on a vev with a Jordan block of size $N$ as explained in \eqref{eq:niltsu}, the quiver becomes 
\begin{equation}\label{eq:ex1magjord}
 {
 1 - 2 - \cdots - N - 2N - 3N -4N -5N-\overset{\overset{\displaystyle 3N}{\vert}}{6N}-4N-2N}\ ,
\end{equation}
corresponding to the massive electric quiver with $n_0=1$
\begin{equation}
\label{eq:ex2}
k-1=N-1=[1^{N-1}]:\ [E_{8}]\ 1\ \overset{\su{1}}{2}\ \overset{\su{2}}{2}\ \overset{\su{3}}{2} \cdots \ \overset{\su{N-1}}{2} \ [\SU(N)]\ ,
\end{equation} 
in agreement with the brane analysis of figure \ref{fig:jordan1}.

On the other hand,  if we turn on a Jordan block of size $N+1$ we get
\begin{equation}\label{eq:ex1magjordbad}
 {
 1 - 2 - \cdots - (N-1) - 2N - 3N -4N -5N-\overset{\overset{\displaystyle 3N}{\vert}}{6N}-4N-2N}\ .
\end{equation}
This is a bad quiver in the sense of \cite{Gaiotto:2008ak}, i.e.  it includes monopole operators which violate the unitarity bound, and its Coulomb branch is not a hyperkähler cone. This issue reflects the fact that in the brane system it is not possible to satisfy the S-rule because the number of D6's ending on the D8 is larger than the number of NS5's.  We will henceforth discard bad quivers and interpret their appearance as evidence of the fact that the vev we are trying to turn on is incompatible with the chiral ring relations defining the Higgs branch of the 6d SCFT. 

\subsection{\texorpdfstring{Electric quivers with $\mathfrak{usp}$ on the $-1$ curve}{Electric quivers with usp on the -1 curve}}

Let us move on to theories with the $-1$ curve supporting a $\mathfrak{usp}$ gauge algebra. This means there are D6-branes crossing the O8$^-$ in the massive IIA setup. For simplicity we first focus on the following example with $n_0=2$,
\begin{equation}
\label{eq:ex3}
k-2=10-2=8=[2'^2,2^2]:\ [E_7]\ \overset{\usp{4}}{1}\ \overset{\su{6}}{2}\ \overset{\su{8}}{2}\ [\SU(10)]\ ,
\end{equation} 
and generalize the analysis later. The corresponding brane system is 
\begin{equation}
\label{eq:brane3}
\begin{tikzpicture}[scale=1,baseline]
\node at (0,0) {};
\draw[fill=black] (1,0) circle (0.075cm);
\draw[fill=black] (2,0) circle (0.075cm);
\draw[fill=black] (3,0) circle (0.075cm);


\draw[solid,black,thick] (0,0)--(1,0) node[black,midway,yshift=0.2cm] {\footnotesize $4$};
\draw[solid,black,thick] (1,0)--(2,0) node[black,midway,yshift=0.2cm] {\footnotesize $6$};
\draw[solid,black,thick] (2,0)--(3,0) node[black,midway,yshift=0.2cm] {\footnotesize $8$};
\draw[solid,black,thick] (3,0)--(4,0) node[black,midway,xshift=0cm,yshift=0.2cm] {\footnotesize $10$};

\draw[dashed,black,very thick] (0,-.5)--(0,.5) node[black,midway, xshift =0cm, yshift=-1.5cm] {} node[black,midway, xshift =0cm, yshift=1.5cm] {};
\draw[solid,black,very thick] (0.25,-.5)--(0.25,.5) node[black,midway, xshift =0cm, yshift=+.75cm] {\footnotesize $6$};
\end{tikzpicture}
\end{equation}
and the magnetic quiver reads
\begin{equation}\label{eq:ex4quiv}
 {
 1 - 2 - \cdots - 9 - 10 - 13 -16 -19-\overset{\overset{\displaystyle 11}{\vert}}{22}-14-6}\ .
\end{equation}
If we turn on a Jordan block of size $N=2$ ($N=3$), we simply bring in another D8 in the above brane system or, equivalently, in the magnetic quiver we replace the $\U(9)$ node with $\U(8)$ ($\U(7)$). This comes from the modification of the $T(\SU(10))$ tail in \eqref{eq:ex4quiv} according to the rule \eqref{eq:niltsu}. A bit more subtle is the case of a Jordan block of size $N=4$, which Higgses the $\mathfrak{usp}$ node in the 6d electric quiver. The effect of the vev is most easily understood at the level of the magnetic quiver, which becomes
\begin{equation}\label{eq:ex5quiv}
 {
 1 - 2 - \cdots - 5- 6 - 10 - 13 -16 -19-\overset{\overset{\displaystyle 11}{\vert}}{22}-14-6}\ .
\end{equation} 
Here we come across a new effect: the $\U(10)$ node is \emph{underbalanced}. In fact at that node we have $2\cdot 10 -1 = 13+6$, i.e. $N_\text{f} = 2N_\text{c}-1$.  Thanks to this fact we can exploit the known 3d duality \cite{Gaiotto:2008ak} between $\U(N)$ SQCD with $N_\text{f} = 2N-1$ and $\U(N-1)$ SQCD with $N_\text{f} = 2N-1$ plus one free hypermultiplet.  By dualizing in such a way all underbalanced nodes, we end up with $12$ free hypermultiplets together with the quiver 
\begin{equation}\label{eq:ex6quiv}
 {
 1 - 2 - \cdots - 5- 6 - 8 - 11 -14 -17-\overset{\overset{\displaystyle 10}{\vert}}{20}-13-6}\ 
\end{equation}
which describes the Higgs branch of the SCFT associated engineered by the brane system 
\begin{equation}
\label{eq:brane4}
\begin{tikzpicture}[scale=1,baseline]
\node at (0,0) {};
\draw[fill=black] (1,0) circle (0.075cm);
\draw[fill=black] (2,0) circle (0.075cm);
\draw[fill=black] (3,0) circle (0.075cm);


\draw[solid,black,thick] (0,0)--(1,0) node[black,midway,yshift=0.2cm] {\footnotesize $2$};
\draw[solid,black,thick] (1,0)--(2,0) node[black,midway,yshift=0.2cm] {\footnotesize $4$};
\draw[solid,black,thick] (2,0)--(3,0) node[black,midway,yshift=0.2cm] {\footnotesize $5$};
\draw[solid,black,thick] (3,0)--(4,0) node[black,midway,xshift=0cm,yshift=0.2cm] {\footnotesize $6$};

\draw[dashed,black,very thick] (0,-.5)--(0,.5) node[black,midway, xshift =0cm, yshift=-1.5cm] {} node[black,midway, xshift =0cm, yshift=1.5cm] {};
\draw[solid,black,very thick] (0.25,-.5)--(0.25,.5) node[black,midway, xshift =0cm, yshift=+.75cm] {\footnotesize $6$};
\draw[solid,black,very thick] (1+0.3,-.5)--(1+0.3,.5) node[black,midway, xshift =0cm, yshift=+.75cm] {\footnotesize $1$};
\end{tikzpicture}\ .
\end{equation}
This result can be also be neatly understood at the level of brane setup.  Starting from the brane system \eqref{eq:brane3} and performing Hanay--Witten moves we find 
\begin{equation}
\label{eq:brane4bis}
\begin{tikzpicture}[scale=1,baseline]
\node at (0,0) {};
\draw[fill=black] (1,0) circle (0.075cm);
\draw[fill=black] (2,0) circle (0.075cm);
\draw[fill=black] (3,0) circle (0.075cm);


\draw[solid,black,thick] (0,0)--(1,0) node[black,midway,yshift=0.2cm] {\footnotesize $3$};
\draw[solid,black,thick] (0,0.4)--(0.7,0.4) node[black,midway,xshift=0.15cm,yshift=0.2cm] {\footnotesize $1$};
\draw[solid,black,thick] (1,0)--(2,0) node[black,midway,yshift=0.2cm] {\footnotesize $4$};
\draw[solid,black,thick] (2,0)--(3,0) node[black,midway,yshift=0.2cm] {\footnotesize $5$};
\draw[solid,black,thick] (3,0)--(4,0) node[black,midway,xshift=0cm,yshift=0.2cm] {\footnotesize $6$};

\draw[dashed,black,very thick] (0,-.5)--(0,.7) node[black,midway, xshift =0cm, yshift=-1.5cm] {} node[black,midway, xshift =0cm, yshift=1.5cm] {};
\draw[solid,black,very thick] (0.25,-.5)--(0.25,.7) node[black,midway, xshift =0cm, yshift=0.85cm] {\footnotesize $6$};
\draw[solid,black,very thick] (0.25+0.45,-.5)--(0.25+0.45,.7) node[black,midway, xshift =0cm, yshift=0.85cm] {\footnotesize $1$};
\end{tikzpicture}\ .
\end{equation} 
The left part of the brane system can be rearranged as follows: 
\begin{equation}
\begin{tikzpicture}
 \draw[D8] (2.5,-.75)--(2.5,.75)  (2.6,-.75)--(2.6,.75)  (2.7,-.75)--(2.7,.75)  (2.8,-.75)--(2.8,.75) (2.9,-.75)--(2.9,.75)  (3,-.75)--(3,.75)  (3.5,-.75)--(3.5,.75);
   \draw[O8]  (2.2,.75)--(2.2,-.75);
   \draw[D6] (2.2,0.5)--(3.5,0.5);
    \node[dot] (a) at (4,0) {};
    \node[] (b) at (5,0) {};
   \draw[D6] (2.2,0)--(a);
   \draw[D6] (a)--(b); 
   \node[] at (3.3,-0.3) {$3$}; 
   \node[] at (4.5,-0.3) {$4$};
   \node[] at (2.2,1.3) {};
   
   \draw[solid, black, thick, ->] (5.5,0)--(6.5,0); 
   
  \draw[D8] (7.5,-.75)--(7.5,.75)  (7.7,-.75)--(7.7,.75)  (7.9,-.75)--(7.9,.75)  (8.1,-.75)--(8.1,.75) (8.3,-.75)--(8.3,.75)  (8.5,-.75)--(8.5,.75)  (8.7,-.75)--(8.7,.75);
   \draw[O8]  (7.2,.75)--(7.2,-.75);
   \draw[D6] (7.3,0.5)--(7.5,0.5) (7.3,0.3)--(7.7,0.3) (7.5,0.6)--(7.7,0.6) (7.7,0.4)--(7.9,0.4) (7.7,0.7)--(7.9,0.7) (7.9,0.5)--(8.1,0.5) (7.9,0.6)--(8.1,0.6) (8.1,0.2)--(8.3,0.2) (8.1,0.7)--(8.3,0.7) (8.3,0.3)--(8.5,0.3) (8.3,0.4)--(8.5,0.4) (8.5,0.2)--(8.7,0.2) (8.5,0.5)--(8.7,0.5);
   \draw[D6] (8.7,0.05)--(9.4+0.075,0.05);
    \node[dot] (a) at (9.5,0) {};
    \node[] (b) at (10.5,0) {};
   \draw[D6] (7.2,0)--(a);
   \draw[D6] (a)--(b); 
   \node[] at (9,-0.3) {$2$}; 
   \node[] at (10,-0.3) {$4$};
   \node[] at (7.2,1.3) {};     
   \draw[D6] (7.3,0.5) arc[start angle=90, end angle=270, radius=0.1cm];
   
     \draw[solid, black, thick, ->] (10.8,0)--(11.8,0); 
     
  \draw[D8] (12.5,-.75)--(12.5,.75)  (12.6,-.75)--(12.6,.75)  (12.7,-.75)--(12.7,.75)  (12.8,-.75)--(12.8,.75) (12.9,-.75)--(12.9,.75)  (13,-.75)--(13,.75)  (14.5,-.75)--(14.5,.75);
   \draw[O8]  (12.2,.75)--(12.2,-.75);
    \node[dot] (a) at (14,0) {};
    \node[] (b) at (15,0) {};
   \draw[D6] (12.2,0)--(a);
   \draw[D6] (a)--(b); 
   \node[] at (13.5,-0.3) {$2$}; 
   \node[] at (14.2,-0.3) {$4$};
   \node[] at (12.2,1.3) {};
\end{tikzpicture}    \ .
\end{equation} 
We first break the D6-branes into segments suspended between D8's. Each segment provides a free hypermultiplet, for a total of twelve, reproducing the result of the magnetic quiver analysis. Once we have remove the D6 segments (i.e. we slide them off to infinity along $x^{789}$ (see e.g. \cite[p.13]{Fazzi:2022hal}), we can bring the last D8 across the NS5 with a Hanany--Witten move. In the process the $\usp{4}$ algebra is Higgsed down to $\usp{2}$. 

By repeating this procedure for general $n_0$ we find $16-2n_0$ free hypermultiplets and the $\usp{2r}$ gauge group gets broken to $\usp{2r-2}$. If we attempt to turn on a bigger Jordan block we find that the magnetic quiver becomes bad. We interpret this as evidence that the Higgsing is not possible due to the chiral ring relations defining the 6d Higgs branch. 

\subsection{\texorpdfstring{Electric quivers with $\mathfrak{su}$ on the $-1$ curve}{Electric quivers with su on the -1 curve}}

Finally let us turn to the case with a half-NS5 stuck on the O8$^-$ plane.  This case corresponds to an $\mathfrak{su}$ gauge algebra with an antisymmetric hypermultiplet supported on the $-1$ curve.  For $n_0=3$ the associated brane system is of the form
\begin{equation}
\label{brane5}
\begin{tikzpicture}
 \draw[D8] (2.5,-.75)--(2.5,.75)  (2.6,-.75)--(2.6,.75)  (2.7,-.75)--(2.7,.75)  (2.8,-.75)--(2.8,.75) (2.9,-.75)--(2.9,.75);
   \draw[O8]  (2.2,.75)--(2.2,-.75);
    \node[dot] (a) at (4,0) {};
    \node[dot] (b) at (5,0) {};
    \node[dot] (c) at (6,0) {};
    \node[dot] (d) at (2.2,0.3) {}; 
    
    \node[] (f) at (7,0) {};
   \draw[D6] (2.2,0)--(a);
   \draw[D6] (a)--(b)--(c)--(f); 
   \node[] at (3.5,-0.3) {$4$}; 
   \node[] at (4.5,-0.3) {$7$};
   \node[] at (5.5,-0.3) {$10$};
   \node[] at (6.5,-0.3) {$13$};
   \node[] at (2.2,1.3) {};
\end{tikzpicture}    \ .
\end{equation} 
In this case a Jordan block of size $N=4$ (the number of full NS5-branes plus one, in general) requires us to let one of the D6's end on the half-NS5. Again, bigger Jordan blocks are not allowed due to the S-rule. Let us analyze in detail these statements via the associated magnetic quivers, which for \eqref{brane5} reads 
\begin{equation}
\label{eq:quiver7} 
 {
 1 - 2 - \cdots - 11 - 12 - 13 -16 -19-\overset{\overset{\displaystyle 11}{\vert}}{22}-14-7}\ .
\end{equation} 
If we turn on a nilpotent vev involving a Jordan block of size $N=4$ we should replace the $\U(12)$ node in \eqref{eq:quiver7} with a $\U(9)$ and accordingly decrease the rank of all nodes on its left by four units. As a result the $\U(13)$ node becomes ugly and can be dualized as above. Once we have removed all ugly nodes we find 
\begin{equation}
\label{eq:quiver8} 
 {
 1 - 2 - \cdots - 8 - 9 - 12 -15 -18-\overset{\overset{\displaystyle 10}{\vert}}{21}-14-7}\ ,
\end{equation} 
accompanied by 5 free hypermultiplets. We can understand how the free hypermultiplets arise by looking again at the brane system: 
\begin{equation}
\begin{tikzpicture}
 \draw[D8] (2.5,-.75)--(2.5,.75)  (2.6,-.75)--(2.6,.75)  (2.7,-.75)--(2.7,.75)  (2.8,-.75)--(2.8,.75) (2.9,-.75)--(2.9,.75) (3.5,-.75)--(3.5,.75);
   \draw[O8]  (2.2,.75)--(2.2,-.75);
   \draw[D6] (2.2,0.3)--(3.5,0.3);
    \node[dot] (a) at (4,0) {};
    \node[dot] (b) at (5,0) {};
    \node[dot] (c) at (6,0) {}; 
    \node[dot] (d) at (2.2,0.3) {}; 
    \node[] (f) at (7,0) {};

   \draw[D6] (2.2,0)--(a);
   \draw[D6] (a)--(b)--(c)--(f); 
   \node[] at (3.2,-0.3) {$3$}; 
   \node[] at (4.5,-0.3) {$5$};
   \node[] at (5.5,-0.3) {$7$};
   \node[] at (6.5,-0.3) {$9$};
   \node[] at (2.2,1.3) {};
\end{tikzpicture}    \ .
\end{equation} 
The left part of the brane system can be rearranged by suspending D6-branes between D8's: 
\begin{equation}
\begin{tikzpicture}
 \draw[D8] (2.5,-.75)--(2.5,.75)  (2.6,-.75)--(2.6,.75)  (2.7,-.75)--(2.7,.75)  (2.8,-.75)--(2.8,.75) (2.9,-.75)--(2.9,.75)  (3.5,-.75)--(3.5,.75);
   \draw[O8]  (2.2,.75)--(2.2,-.75);
   \draw[D6] (2.3-0.075,0.3)--(3.5,0.3);
    \node[dot] (a) at (4,0) {};
    \node[dot] (d) at (2.2,0.3) {}; 
    \node[] (b) at (5,0) {};
   
   \draw[D6] (2.2,0)--(a);
   \draw[D6] (a)--(b); 
   \node[] at (3.2,-0.3) {$3$}; 
   \node[] at (4.5,-0.3) {$5$};
   \node[] at (2.2,1.3) {};
   
   \draw[D6,->] (5.5,0)--(6.5,0); 
   
  \draw[D8] (7.5,-.75)--(7.5,.75)  (7.7,-.75)--(7.7,.75)  (7.9,-.75)--(7.9,.75)  (8.1,-.75)--(8.1,.75) (8.3,-.75)--(8.3,.75)  (8.5,-.75)--(8.5,.75);
   \draw[O8]  (7.2,.75)--(7.2,-.75);
   \draw[D6] (7.3-0.075,0.3)--(7.5,0.3) (7.5,0.4)--(7.7,0.4) (7.7,0.5)--(7.9,0.5) (7.9,0.6)--(8.1,0.6) (8.1,0.4)--(8.3,0.4) (8.3,0.3)--(8.5,0.3);
    \node[dot] (a) at (9.5,0) {};
    \node[] (b) at (10.5,0) {};
     \node[dot] (d) at (7.2,0.3) {}; 
   
   \draw[D6] (7.2,0)--(a);
   \draw[D6] (a)--(b); 
   \node[] at (9,-0.3) {$3$}; 
   \node[] at (10,-0.3) {$5$};
   \node[] at (7.2,1.3) {};     
   
	\draw[D6, ->] (10.8,0)--(11.8,0);    
  \draw[D8] (12.5,-.75)--(12.5,.75)  (12.6,-.75)--(12.6,.75)  (12.7,-.75)--(12.7,.75)  (12.8,-.75)--(12.8,.75) (12.9,-.75)--(12.9,.75)  (13,-.75)--(13,.75);
   \draw[O8]  (12.2,.75)--(12.2,-.75);
    \node[dot] (a) at (14,0) {};
    \node[] (b) at (15,0) {};
     \node[dot] (d) at (12.2,0.3) {};
   
   \draw[D6] (12.2,0)--(a);
   \draw[D6] (a)--(b); 
   \node[] at (13.5,-0.3) {$3$}; 
   \node[] at (14.5,-0.3) {$5$};
   \node[] at (12.2,1.3) {};
\end{tikzpicture}    
\end{equation} 
where in the last step we have removed the 5 D6 segments suspended between D8's, each providing a free hypermultiplet, and then we have moved the leftmost D8 through the orientifold plane: due to the Hanany--Witten effect the brane suspended between it and the half-NS5 disappears as it crosses the O8$^-$-plane. At the same time the image D8 enters and becomes the new leftmost brane. In short we have the transition $n_0\to n_0-1$ accompanied by the production of $n_0$ free hypermultiplets. 

There is one exception to this rule when the gauge algebra supported on the $-1$ curve is broken all the way to $\su{2}$. In this case we have the production of $n_0+1$ hypermultiplets. The origin of the extra free field is easy to understand.  In the Higgsed theory the half-NS5 contributes a hypermultiplet in the antisymmetric of $\mathfrak{su}$ and this coincides with the trivial representation when that algebra is $\su{2}$. This observation is indeed reproduced by the magnetic quiver analysis. If we consider \eqref{eq:quiver8}, which describes a SCFT whose $-1$ curve supports $\su{3}$, and turn on again a Jordan block of size $N=4$, we obtain the quiver 
\begin{equation}
\label{eq:quiver9} 
 {
 1 - 2 - \cdots - 5 - 8- 11 -14 -17-\overset{\overset{\displaystyle 10}{\vert}}{20}-13-6}\ .
\end{equation}
This includes as usual underbalanced nodes which produce free hypermultiplets via dualization. Overall, we find 6 hypermultiplets as expected, plus the magnetic quiver 
\begin{equation}
\label{eq:quiver10} 
 {
 1 - 2 - \cdots - 5 - 8 - 11 -14 -17-\overset{\overset{\displaystyle 10}{\vert}}{20}-13-6}\ .
\end{equation} 
which describes the brane system 
\begin{equation}
\begin{tikzpicture}
 \draw[D8] (2.5,-.75)--(2.5,.75)  (2.6,-.75)--(2.6,.75)  (2.7,-.75)--(2.7,.75)  (2.8,-.75)--(2.8,.75) (2.9,-.75)--(2.9,.75) (3.0,-.75)--(3.0,.75) (3.1,-.75)--(3.1,.75);
   \draw[O8]  (2.2,.75)--(2.2,-.75);
    \node[dot] (a) at (4,0) {};
    \node[dot] (b) at (5,0) {};
    \node[dot] (c) at (6,0) {}; 
    \node[dot] (d) at (2.2,0.3) {}; 
    \node[] (f) at (7,0) {};
   
   \draw[D6] (2.2,0)--(a);
   \draw[D6] (a)--(b)--(c)--(f); 
   \node[] at (3.5,-0.3) {$2$}; 
   \node[] at (4.5,-0.3) {$3$};
   \node[] at (5.5,-0.3) {$4$};
   \node[] at (6.5,-0.3) {$5$};
   \node[] at (2.2,1.3) {};
\end{tikzpicture}    \ .
\end{equation} 
Lastly, if we turn on again a nilpotent vev in the resulting theory we find in the IR the rank-$3$ E-string theory accompanied by the usual free sector. 

All we have done so far is not directly applicable to $n_0>7$ since the magnetic quiver does not have the structure described before, with a $T(\SU(k))$ theory coupled to another quiver. For large $n_0$ the balanced subquiver responsible for the $\SU(k)$ symmetry includes the central node and can rather be described by a $T_{\rho}^{\sigma}(\SU(k))$ theory \cite{Cremonesi:2014uva}, with $\rho=(1^k)$, so only the $\sigma$ partition is nontrivial. Let us discuss one example to illustrate this point. 

Consider the theory with $n_0=9$ for $k=24$. The gauge algebras are $\su{6}$, supported on the $-1$ curve, and $\su{15}$. The corresponding magnetic quiver is 
\begin{equation}
\label{eq:quiver9bis} 
 {
 1 - 2 - 3- \cdots - 17- 18 -19 -20-\overset{\overset{\displaystyle 8}{\vert}}{21}-14-7}\ ,
\end{equation} 
and all the nodes are balanced except the $\U(8)$ node on top. If we ungauge it, we obtain the theory $T_{\rho}^{\sigma}(\SU(24))$ with $\rho=(1^{24})$ and $\sigma=(3^8)$. 

The idea now is to describe the effect of the nilpotent vev for the $\SU(24)$ symmetry by changing $\rho$ and leaving $\sigma$ untouched. Finally, we regauge the $\U(8)$ node to obtain the magnetic quiver of the IR SCFT. If we want for example to consider a Jordan block of size 2 for the theory at hand we should replace $\rho=(1^{24})$ with $\rho=(2,1^{22})$. Once we have regauged  the $\U(8)$ node we land on  
\begin{equation}
\label{eq:quiver10bis} 
 {
 1 - 2 - 3- \cdots - 17- 18 -19 -\overset{\overset{\displaystyle 8}{\vert}}{20}-13-6}\ ,
\end{equation}
which we claim to be the magnetic quiver associated with the IR SCFT. As a consistency check, we can notice that, upon by subtracting the $E_8$ quiver twice, \eqref{eq:quiver10bis} becomes the mirror dual of a gauge theory with gauge groups $\SU(6)$ and $\SU(14)$. We therefore see that the $\SU(15)$ gauge group is Higgsed to $\SU(14)$, in agreement with the brane intuition.

\section{Conclusions}
\label{sec:conc}

This work complements naturally what done in \cite{Giacomelli:2022drw,Fazzi:2022hal} for orbi-instantons, and extends the techniques used there to construct the hierarchy of Higgs branch RG flows to the infinite class of 6d SCFTs known as massive E-string theories. We have seen that, on top of flows between massive theories defined by different constrained Kac labels at fixed $n_0$, there also exist flows involving the right $\SU(k)$ flavor symmetry factor (which were neglected for orbi-instantons for clarity of exposition, even though they \emph{are} part of their \emph{full} Higgs branch). These flows allow to go from a massive theory defined by $n_0$ to one by a different $n_0$ upon switching on nontrivial Jordan blocks in the vev of hypermultiplets charged under $\SU(k)$.  We have seen that these flows actually ``mix'' left and right flavor symmetry factors of the SCFT, since these Jordan blocks have the nontrivial effect of propagating an unbalancing from the left to the right tail of the 3d magnetic quiver. Upon rebalancing, we land on a new massive theory.

An obvious extension of this work was already mentioned in the conclusions of \cite{Fazzi:2022hal}.  Here we have found that the $E_8$ Kac labels, constrained as per table \ref{tab:constrE8}, correspond to massive E-strings with electric quiver as in \eqref{eq:massive} (or appropriate generalizations with nonempty $-1$ curve, see e.g. figure \ref{fig:E8-IIA}).  However one may also ask whether the $E_{1+(8-n_0)}$ Kac labels (which can be defined in the obvious way, using the Coxeter labels of the affine version of $E_{1+(8-n_0)}$ for $n_0>1$ \cite{kac1990infinite}) play any role in the classification of (massive) 6d SCFTs. Simple counting arguments show that there exist many more $E_{1+(8-n_0)}$ Kac labels than constrained $E_8$ Kac labels,  so one could ask whether the former correspond to new 6d SCFTs. By the same token, we could also consider the \emph{twisted} affine $E_{1+(8-n_0)}$ Dynkin diagrams (which are obtained by acting with an outer automorphism on the extended Dynkin, see \cite[Tab. 4]{Fazzi:2022hal}) to construct even more Kac labels.\footnote{The twisted affine Dynkin diagrams are obtained by folding the corresponding untwisted ones, see \cite[Chap. 30]{Bump2013-gy} (and the original reference \cite{kac1990infinite}) for root foldings in a Lie algebra. The concept has already found applications in 3d \cite{Gulotta:2012yd}, 2d CFTs \cite{Fuchs:1995zr, DiFrancesco:1997nk}, and much more recently in 4d $\mathcal{N}=3$ SCFTs \cite{Giacomelli:2020gee}.  A possible obstacle in 6d constructions is the absence of an obvious ingredient in perturbative Type IIA which implements the $\zz_2$ automorphism (which is order-two for the Lie algebras in the $E_{1+(8-n_0)}$ list). This could come from an additional $\zz_2$ orbifold.} It would be interesting to understand whether these too play a role in the classification of massive E-strings, and whether one can construct flows for these by subtracting the associated magnetic quivers (perhaps by using the techniques of \cite{Bourget:2020bxh}). If this turns out to be correct, it may suggest the existence of small $E_{1+(8-n_0)}$ instanton transition beyond the well-known $E_8$ case (in which one tensor multiplet is transmuted into twenty-nine hypermultiplets).  A different but related question which was already asked in section \ref{subsub:5d} is whether the massive E-strings give rise to new high-rank 5d SCFTs.

Another generalization comes from considering D-type orbi-instantons rather than A-type to construct the associated massive E-strings (as done here around \eqref{eq:massEntail}-\eqref{eq:massive}). The Type IIA picture contains a combined O8-O6 projection,\footnote{And at their intersection a so-called ON$^0$ as well, see \cite[footnote 40]{Fazzi:2022hal}.} and some of these massive theories have already been constructed in \cite[Sec. 3.4]{Apruzzi:2017nck}. A different possible generalization involves the O8$^0$ and O8$^{-1}$ planes of \cite{Keurentjes:2000bs,Aharony:2007du}. One could ask whether adding some D8's on top of either of the two produces another class of massive E-strings, which could be amenable to be studied (together with their flows) by the same techniques adopted here.

Finally, the most important conjecture we can formulate \cite{Fazzi:2022hal,fazzi-giri-levy} is that the hierarchies of RG flows of orbi-instantons are nothing but Hasse diagrams of $E_{8}$-orbits of the \emph{double} affine Grassmannian of $E_{8}$ \cite{Braverman:2007dvq}.  The double in the name is related to the fact that one uses the affine $E_8$ Dynkin to construct (via dominant coweights) the Grassmannian (see e.g. \cite{Bourget:2021siw} for an introduction aimed at physicists), rather than the finite one (as in the singly affine Grassmannian). The number $N$ of M5's/NS5's is related to the imaginary roots in the affine $E_8$ Kac--Moody algebra (see e.g.  \cite[Chap. 12]{fuchs2003symmetries}).  Slicing the Hasse along a subdiagram at fixed $N$ produces the graphs in \cite[App. A]{Fazzi:2022hal}; considering the full Hasse produces instead a hierarchy which starts in $E_8$ at $N$ and ends in $E_8$ at $N-1$ (all the way down to $N=1$), i.e. we are performing small instanton transitions.  (See e.g. \cite[Fig. 12]{Fazzi:2022hal}, which is an early result for $k=4$ from \cite{Frey:2018vpw}, or the $k$'s considered in \cite{Giacomelli:2022drw}.) Then, for massive E-strings the hierarchies produced in section \ref{sec:higgsflows} must be understood as disconnected subdiagrams of the full Hasse albeit for a nongeneric orbi-instanton at $k-n_0$ with length-one plateau, where only some flows are considered (those between constrained Kac labels of $k$, given the choice of $n_0$). From this perspective, it is also clear the examples of section \ref{sec:Tflows} provide a perspective on a yet different ``wedge'' of the full Higgs branch of those massive strings, since there we are activating vev's for the right $\SU(k)$ factor which propagate to the far left, landing us on a massive string at a different value of $n_0$ with respect to the one we started with.

The structure of the Higgs branch of 6d $(1,0)$ theories is obviously extremely complicated.  The above conjecture (if proven) would open a completely new window into the study of 6d Higgs branches at infinite coupling, including their \emph{geometry}. This is important, as \cite{Mekareeya:2017jgc} has identified the Higgs branch of orbi-instantons with the moduli space of $N$ $E_8$ instantons on an ALE space (the orbifold), for which there exists no ADHM construction (although see \cite{Mekareeya:2015bla}). We plan to come back to this point in the future.

\section*{Acknowledgments}

We would like to thank Fabio Apruzzi, Oren Bergman, Antoine Bourget, Julius Grimminger,  Paul Levy, Noppadol Mekareeya, Alessandro Tomasiello, Gabi Zafrir for useful comments and discussions.  MF would like to thank the University of Milano--Bicocca for the warm hospitality in the last stages of this work.  MF and Simone Giacomelli gratefully acknowledge support from the Simons Center for Geometry and Physics (workshops ``5d $\mathcal{N}=1$ SCFTs and Gauge Theories on Brane Webs'' and ``Supersymmetric Black Holes, Holography and Microstate Counting''), Stony Brook University at which some of the research for this paper was performed.  The work of MF is supported in part by the Knut and Alice Wallenberg Foundation under grant KAW 2021.0170, the VR grant 2018-04438, the Olle Engkvists Stiftelse grant No. 2180108,  the European Union's Horizon 2020 research and innovation programme under the Marie Skłodowska-Curie grant agreement No. 754496 - FELLINI  and the European Research Council (ERC) grant agreement No. 851931.  The work of Simone Giacomelli is supported by the INFN grant ``Per attività di formazione per sostenere progetti di ricerca'' (GRANT 73/STRONGQFT). The work of Suvendu Giri is supported in part by INFN and by MIUR-PRIN contract 2017CC72MK003.

\appendix

\section{Rules for magnetic quivers}
\label{app:rules}

In this appendix we list the rules to construct the magnetic quivers associated with the three different classes of 6d electric quivers we can have, namely those that start out with an empty $-1$ curve, those with $\mathfrak{usp}$ on the $-1$, those with $\mathfrak{su}$ (and one antisymmetric). We will deal with the three classes separately. The magnetic quivers are used to construct allowed RG flows between fixed points via quiver subtraction \cite{Cabrera:2018ann}, as in \cite{Giacomelli:2022drw,Fazzi:2022hal}. In practice, we follow this simple algorithm:
\begin{itemize}
\item choose $n_0$ and $k$;
\item write down all possible constrained Kac labels for $k-n_0$;
\item use the algorithm in \cite{Mekareeya:2017jgc} to construct the associated 6d electric quivers;
\item use the rules given below to construct the 3d magnetic quivers of the 6d electric ones;
\item use quiver subtraction to construct the allowed flows.
\end{itemize}
In the quivers below we will just write $[E_{\ldots}]$ to indicate the left flavor symmetry factor associated with the constrained Kac label, which generically is a subalgebra of $E_{1+(8-n_0)}$.

\subsection{Empty $-1$ curve}

For 6d quivers of the type 
\begin{equation}
[E_{\ldots}]\ 1 \ \overset{\su{r_1}}{\underset{[N_\text{f}=f_1]}{2}} \ \overset{\su{r_2}}{\underset{[N_\text{f}=f_2]}{2}} \ \cdots\ \overset{\su{r_n}}{2}\ [\SU(f_n)]\ .
\end{equation}
The magnetic quiver is given by
\begin{equation}\label{eq:magempty}
\begin{tikzpicture}[baseline]
	\node[] (L1) at (0,0) {$1$};
	\node[] (L2) at (0.75,0) {$2$};
	\node[] (L3) at (1.75,0) {$\dots$};
	\node[] (L4) at (3,0) {$r_n$};
	\node[] (L5) at (5.5,0) {$x_1-x_2-\cdots-x_i$};
	\node[] (G1) at (2.5,0.8) {$1$};
	\node[] (G2) at (3,0.8) {$\cdots$};
	\node[] (G3) at (3.5,0.8) {$1$};
	\node[] (X1) at (3,1.2) {$\overbrace{\phantom{1\cdots1}}^{N}$};
	\draw  (L1) to (L2) to (L3) to (L4) to (L5); 
	\draw (G1) to (L4);
	\draw  (G3) to (L4);
\end{tikzpicture}
\end{equation}
where the $x_i$ in the quiver are given by the sum of all boxes to the right of (not including) the $i$-th column in the following Young tableau (which contains $r_{i+1} - r_i$ boxes in each row). 
\begin{table}[h!]
\centering
	\begin{tabular}{lllllll}
		\cline{2-7}
		\multicolumn{1}{l|}{$r_1 \rightarrow$}           & \multicolumn{1}{l|}{} & \multicolumn{1}{l|}{} & \multicolumn{2}{l|}{$\cdots$} & \multicolumn{1}{l|}{} & \multicolumn{1}{l|}{} \\ \cline{2-7} 
		\multicolumn{1}{l|}{$r_2-r_1 \rightarrow$}       & \multicolumn{1}{l|}{} & \multicolumn{1}{l|}{} & \multicolumn{2}{l|}{$\cdots$} &                       &                       \\ \cline{2-5}
		& $\vdots$              &                       &               &               &                       &                       \\ \cline{2-5}
		\multicolumn{1}{l|}{$r_n - r_{n-1} \rightarrow$} & \multicolumn{1}{l|}{} & \multicolumn{3}{l|}{$\cdots$}                         &                       &                       \\ \cline{2-5}
	\end{tabular}
\end{table}
The actual values of the $x_i$ depend on the chosen Kac label.
As an example, the quiver in \eqref{eq:ex1} 
\begin{equation}
	[E_{7}]\ 1\ \overset{\mathfrak{su}(2)}{2}\ \overset{\mathfrak{su}(4)}{2}\ \overset{\mathfrak{su}(6)}{2} \cdots \ \overset{\mathfrak{su}(2N-2)}{2} \ [\SU(2N)] \ ,
\end{equation}
gives
\begin{equation}
	\begin{tikzpicture}[baseline]
		\node[] (L1) at (0,0) {$1$};
		\node[] (L2) at (0.85,0) {$2$};
		\node[] (L3) at (2,0) {$\dots$};
		\node[] (L4) at (3.75,0) {$(2N-1)$};
		\node[] (L5) at (6,0) {$(2N-2)$};
		\node[] (L6) at (7.5,0) {$2$};
		\node[] (G1) at (5,1) {$1$};
		\node[] (G2) at (6,1) {$\cdots$};
		\node[] (G3) at (6.75,1) {$1$};
		\node[] (X1) at (6,1.5) {$\overbrace{\phantom{\cdots1\cdots\quad}}^{N}$};
		\draw  (L1) to (L2) to (L3) to (L4) to (L5) to (L6); 
		\draw (G1) to (L5);
		\draw  (G3) to (L5);
	\end{tikzpicture}
\end{equation}
After $N$ small instanton transitions (which add $N$ copies of the affine $E₈$ Dynkin diagram), this becomes \eqref{eq:ex1mag}, i.e.
\begin{equation}
	1 - 2 - \cdots - (2N-1) - 2N-3N -4N -5N-\overset{\overset{\displaystyle 3N}{\vert}}{6N}-4N-2N\ .
\end{equation}

\subsection{\texorpdfstring{$\mathfrak{usp}$ on the $-1$ curve}{usp on the -1 curve}}
For quivers of the type
\begin{equation}
[E_{\ldots}]\ \overset{\usp{2r_0}}{\underset{[N_\text{f}=f_0]}{1}} \ \overset{\su{r_1}}{\underset{[N_\text{f}=f_1]}{2}} \ \overset{\su{r_2}}{\underset{[N_\text{f}=f_2]}{2}} \ \cdots\ \overset{\su{r_n}}{2}\ [\SU(f_n)]\ ,
\end{equation}
the magnetic quiver is given by
\begin{equation}\label{eq:magusp}
	\begin{tikzpicture}[baseline]
		\node[] (L1) at (0,0) {$1$};
		\node[] (L2) at (1,0) {$2$};
		\node[] (L3) at (2,0) {$\dots$};
		\node[] (L4) at (3,0) {$r_n$};
		\node[] (L5) at (4,0) {$x_1$};
		\node[] (L56) at (5,0) {$\cdots$};
		\node[] (L6) at (7.2,0) {$2r_0-2r_0-\cdots$};
		\node[] (L7) at (9.5,0) {$2r_0$};
		\node[] (L8) at (10.5,0) {$r_0$};
		\node[] (G1) at (2.5,0.8) {$1$};
		\node[] (G2) at (3,0.8) {$\cdots$};
		\node[] (G3) at (3.5,0.8) {$1$};
		\node[] (G4) at (9.5,0.8) {$r_0$};
		\node[] (X1) at (3,1.2) {$\overbrace{\phantom{1\cdots1}}^{N}$};
		\node[] (X2) at (6.25,-0.5) {$\underbrace{\hskip 200pt}_{\text{7 nodes}}$};
		\draw  (L1) to (L2) to (L3) to (L4) to (L5) to (L56) to (L6) to (L7) to (L8); 
		\draw (G1) to (L4);
		\draw  (G3) to (L4) (G4) to (L7);
	\end{tikzpicture}
\end{equation}
where the tail in \eqref{eq:magempty} is truncated at $x_{i}=2r_0$ and replaced with the new tail such that the trivalent node $\U(2r_0)$ is 7 nodes away from the node $\U(r_n)$. Taking the example of \eqref{eq:ex3} with $N=3, r_0=2, r_n=8$ we get
\begin{equation}
	\begin{tikzpicture}[baseline]
		\node[] (L1) at (0,0) {$1$};
		\node[] (L2) at (1,0) {$2$};
		\node[] (L3) at (2,0) {$\dots$};
		\node[] (L4) at (3,0) {$8$};
		\node[] (L5) at (5.25,0) {$6-4-4-4-4$};
		\node[] (L6) at (7.5,0) {$4$};
		\node[] (L7) at (8.25,0) {$2$};
		\node[] (L8) at (9,0) {$2$};
		\node[] (G1) at (2.5,0.8) {$1$};
		\node[] (G2) at (3,0.8) {$\cdots$};
		\node[] (G3) at (3.5,0.8) {$1$};
		\node[] (G4) at (8.25,0.8) {$2$};
		\node[] (X1) at (3,1.2) {$\overbrace{\phantom{1\cdots1}}^{3}$};
		\draw  (L1) to (L2) to (L3) to (L4) to (L5) to (L6) to (L7) to (L8); 
		\draw (G1) to (L4);
		\draw  (G3) to (L4) (G4) to (L7);
	\end{tikzpicture}
\end{equation}
which after $3$ instanton transitions gives \eqref{eq:ex4quiv},
\begin{equation}
		1 - 2 - \cdots - 9 - 10 - 13 -16 -19-\overset{\overset{\displaystyle 11}{\vert}}{22}-14-6\ .
\end{equation}

\subsection{\texorpdfstring{$\mathfrak{su}$ on the $-1$ curve}{su on the -1 curve}}
The case of a $-1$ curve decorated by an $\su{r}$ algebra can be further subdivided into three subcases:
\begin{enumerate}
	\item[\emph{i)}] $r=2m$ with a two-index antisymmetric hypermultiplet of $\su{2m}$;
	\item[\emph{ii)}] $r=2m+1$ with a two-index antisymmetric hypermultiplet of $\su{2m+1}$;
	\item[\emph{iii)}] $r=6$ with a three-index antisymmetric half-hypermultiplet of $\su{6}$.
\end{enumerate}

\subsubsection{\texorpdfstring{$r=2m$}{r=2m}}
For quivers of the type
\begin{equation}
[E_{\ldots}]\ \overset{\su{2m}}{\underset{[N_{\fontsize{0.5pt}{1pt}\selectfont \yng(1,1)}=1]}{1}} \ \overset{\su{r_1}}{\underset{[N_\text{f}=f_1]}{2}} \ \overset{\su{r_2}}{\underset{[N_\text{f}=f_2]}{2}} \ \cdots\ \overset{\su{r_n}}{2}\ [\SU(f_n)]\ ,
\end{equation}
the magnetic quiver is similar to \eqref{eq:magusp},  except for the tail:
\begin{equation}
	\begin{tikzpicture}[baseline]
		\node[] (L1) at (0,0) {$1$};
		\node[] (L2) at (1,0) {$2$};
		\node[] (L3) at (2,0) {$\dots$};
		\node[] (L4) at (3,0) {$r_n$};
		\node[] (L5) at (4,0) {$x_1$};
		\node[] (L56) at (5,0) {$\cdots$};
		\node[] (L6) at (7.2,0) {$2m-2m-\cdots$};
		\node[] (L7) at (9.5,0) {$2m$};
		\node[] (L8) at (10.5,0) {$m$};
		\node[] (L9) at (11.4,0) {$1$};
		\node[] (G1) at (2.5,0.8) {$1$};
		\node[] (G2) at (3,0.8) {$\cdots$};
		\node[] (G3) at (3.5,0.8) {$1$};
		\node[] (G4) at (9.5,0.8) {$m$};
		\node[] (X1) at (3,1.2) {$\overbrace{\phantom{1\cdots1}}^{N}$};
		\node[] (X2) at (6.25,-0.5) {$\underbrace{\hskip 200pt}_{\text{7 nodes}}$};
		\draw  (L1) to (L2) to (L3) to (L4) to (L5) to (L56) to (L6) to (L7) to (L8) to (L9); 
		\draw (G1) to (L4);
		\draw  (G3) to (L4) (G4) to (L7);
	\end{tikzpicture}\ .
\end{equation}

\subsubsection{\texorpdfstring{$r=2m+1$}{r=2m+1}}
For quivers of the type
\begin{equation}
[E_{\ldots}]\ \overset{\su{2m+1}}{\underset{[N_{\fontsize{0.5pt}{1pt}\selectfont \yng(1,1)}=1]}{1}} \ \overset{\su{r_1}}{\underset{[N_\text{f}=f_1]}{2}} \ \overset{\su{r_2}}{\underset{[N_\text{f}=f_2]}{2}} \ \cdots\ \overset{\su{r_n}}{2}\ [\SU(f_n)]\ ,
\end{equation}
the tail is now given by
\begin{equation}
	\begin{tikzpicture}[baseline]
		\node[] (L1) at (0,0) {$1$};
		\node[] (L2) at (0.75,0) {$2$};
		\node[] (L3) at (2,0) {$\dots$};
		\node[] (L4) at (3,0) {$r_n$};
		\node[] (L5) at (4.5,0) {$x_1-\cdots$};
		\node[] (L6) at (8.25,0) {$(2m+1)-(2m+1)-\cdots$};
		\node[] (L7) at (12,0) {$(2m+1)$};
		\node[] (L8) at (14,0) {$(m+1)$};
		\node[] (L9) at (15.25,0) {$1$};
		\node[] (G1) at (2.5,0.8) {$1$};
		\node[] (G2) at (3,0.8) {$\cdots$};
		\node[] (G3) at (3.5,0.8) {$1$};
		\node[] (G4) at (12,0.8) {$m$};
		\node[] (X1) at (3,1.2) {$\overbrace{\phantom{1\cdots1}}^{N}$};
		\node[] (X2) at (7.5,-0.5) {$\underbrace{\hskip 270pt}_{\text{7 nodes}}$};
		\draw  (L1) to (L2) to (L3) to (L4) to (L5) to (L6) to (L7) to (L8) to (L9); 
		\draw (G1) to (L4);
		\draw  (G3) to (L4) (G4) to (L7);
	\end{tikzpicture}\ .
\end{equation}

\subsubsection{\texorpdfstring{$r=6$}{r=6}}
For quivers of the type
\begin{equation}
[E_{\ldots}]\ \overset{\su{6}}{\underset{[N_{\fontsize{0.5pt}{1pt}\selectfont \frac{1}{2}\yng(1,1,1)}=1]}{1}} \ \overset{\su{r_1}}{\underset{[N_\text{f}=f_1]}{2}} \ \overset{\su{r_2}}{\underset{[N_\text{f}=f_2]}{2}} \ \cdots\ \overset{\su{r_n}}{2}\ [\SU(f_n)]\,,
\end{equation}
the tail becomes
\begin{equation}
	\begin{tikzpicture}[baseline]
		\node[] (L1) at (0,0) {$1$};
		\node[] (L2) at (0.75,0) {$2$};
		\node[] (L3) at (2,0) {$\dots$};
		\node[] (L4) at (3,0) {$r_n$};
		\node[] (L5) at (4.5,0) {$x_1-\cdots$};
		\node[] (L6) at (6.8,0) {$6-6-\cdots$};
		\node[] (L7) at (8.5,0) {$6$};
		\node[] (L8) at (9.25,0) {$4$};
		\node[] (L9) at (10,0) {$2$};
		\node[] (G1) at (2.5,0.8) {$1$};
		\node[] (G2) at (3,0.8) {$\cdots$};
		\node[] (G3) at (3.5,0.8) {$1$};
		\node[] (G4) at (8.5,0.8) {$2$};
		\node[] (X1) at (3,1.2) {$\overbrace{\phantom{1\cdots1}}^{N}$};
		\node[] (X2) at (5.7,-0.5) {$\underbrace{\hskip 170pt}_\text{7 nodes}$};
		\draw  (L1) to (L2) to (L3) to (L4) to (L5) to (L6) to (L7) to (L8) to (L9); 
		\draw (G1) to (L4);
		\draw  (G3) to (L4) (G4) to (L7);
	\end{tikzpicture}\ .
\end{equation}
In writing this, we are assuming that the $\U(6)$ comes before the trivalent node of affine $E_8$, which is indeed the case whenever the $\su{6}$ supported on the $-1$ curve has at least three fundamental flavors. The quiver is instead modified as follows in the other cases. When there are two flavors we have 
\begin{equation}
	\begin{tikzpicture}[baseline]
		\node[] (L1) at (0,0) {$1$};
		\node[] (L2) at (0.75,0) {$2$};
		\node[] (L3) at (2,0) {$\dots$};
		\node[] (L4) at (3,0) {$r_n$};
		\node[] (L5) at (4.5,0) {$x_1-x_2$};
		\node[] (L6) at (6.8,0) {$x_3-x_4-x_5$};
		\node[] (L7) at (8.5,0) {$x_6$};
		\node[] (L8) at (9.25,0) {$4$};
		\node[] (L9) at (10,0) {$2$};
		\node[] (G1) at (2.5,0.8) {$1$};
		\node[] (G2) at (3,0.8) {$\cdots$};
		\node[] (G3) at (3.5,0.8) {$1$};
		\node[] (G4) at (8.5,0.8) {$2$};
		\node[] (X1) at (3,1.2) {$\overbrace{\phantom{1\cdots1}}^{N}$};
		\node[] (X2) at (5.7,-0.5) {$\underbrace{\hskip 170pt}_\text{7 nodes}$};
		\draw  (L1) to (L2) to (L3) to (L4) to (L5) to (L6) to (L7) to (L8) to (L9); 
		\draw (G1) to (L4);
		\draw  (G3) to (L4) (G4) to (L7);
	\end{tikzpicture}\ .
\end{equation}
If we have only one flavor we should use 
\begin{equation}
	\begin{tikzpicture}[baseline]
		\node[] (L1) at (0,0) {$1$};
		\node[] (L2) at (0.75,0) {$2$};
		\node[] (L3) at (2,0) {$\dots$};
		\node[] (L4) at (3,0) {$r_n$};
		\node[] (L5) at (4.5,0) {$(x_1-x_2)$};
		\node[] (L6) at (7,0) {$(x_3-x_4-x_5)$};
		\node[] (L7) at (9,0) {$x_6$};
		\node[] (L8) at (10.5,0) {$(x_7-2)$};
		\node[] (L9) at (11.75,0) {$2$};
		\node[] (G1) at (2.5,0.8) {$1$};
		\node[] (G2) at (3,0.8) {$\cdots$};
		\node[] (G3) at (3.5,0.8) {$1$};
		\node[] (G4) at (9,0.8) {$2$};
		\node[] (X1) at (3,1.2) {$\overbrace{\phantom{1\cdots1}}^{N}$};
		\node[] (X2) at (5.7,-0.5) {$\underbrace{\hskip 170pt}_\text{7 nodes}$};
		\draw  (L1) to (L2) to (L3) to (L4) to (L5) to (L6) to (L7) to (L8) to (L9); 
		\draw (G1) to (L4);
		\draw  (G3) to (L4) (G4) to (L7);
	\end{tikzpicture}\ .
\end{equation} 
Finally, in the case without flavors we have 
\begin{equation}
	\begin{tikzpicture}[baseline]
		\node[] (L1) at (0,0) {$1$};
		\node[] (L2) at (0.75,0) {$2$};
		\node[] (L3) at (2,0) {$\dots$};
		\node[] (L4) at (3,0) {$r_n$};
		\node[] (L5) at (4.5,0) {$(x_1-x_2)$};
		\node[] (L6) at (7,0) {$(x_3-x_4-x_5)$};
		\node[] (L7) at (9,0) {$x_6$};
		\node[] (L8) at (10.5,0) {$(x_7-2)$};
		\node[] (L9) at (12.5,0) {$(x_8-4)$};
		\node[] (G1) at (2.5,0.8) {$1$};
		\node[] (G2) at (3,0.8) {$\cdots$};
		\node[] (G3) at (3.5,0.8) {$1$};
		\node[] (G4) at (9,0.8) {$2$};
		\node[] (X1) at (3,1.2) {$\overbrace{\phantom{1\cdots1}}^{N}$};
		\node[] (X2) at (5.7,-0.5) {$\underbrace{\hskip 170pt}_\text{7 nodes}$};
		\draw  (L1) to (L2) to (L3) to (L4) to (L5) to (L6) to (L7) to (L8) to (L9); 
		\draw (G1) to (L4);
		\draw  (G3) to (L4) (G4) to (L7);
	\end{tikzpicture}\ .
\end{equation}


\bibliography{main}
\bibliographystyle{at}

\end{document}